\documentclass[draftclsnofoot,onecolumn,12pt]{IEEEtran}

\ifCLASSINFOpdf
\else
\fi

\usepackage{setspace}
\usepackage{bbm}
\usepackage[cmex10]{amsmath}
\usepackage{amssymb}
\usepackage{cite}
\usepackage{graphicx}
\usepackage{array,color}
\usepackage{amsmath}
\allowdisplaybreaks
\usepackage{stfloats}
\usepackage{graphicx}
\usepackage{epstopdf}
\usepackage{subfigure}
\usepackage{tabularx}
\usepackage{epsfig,epsf,color,balance,cite}
\usepackage{algorithmic}
\usepackage{algorithm}
\usepackage{url}
\usepackage{bm}
\usepackage{multirow}

\usepackage{amsthm}

\ifodd 0
\usepackage{soul}
\usepackage{color}
\setstcolor{red}

\newcommand{\rev}[1]{{\color{red}#1}} 
\newcommand{\del}[1]{\st{#1}} 

\newcommand{\com}[1]{\textbf{\color{red} (COMMENT: #1)}} 
\newcommand{\response}[1]{\textbf{\color{green} (RESPONSE: #1)}} 
\else

\newcommand{\rev}[1]{#1}

\newcommand{\del}[1]{}

\newcommand{\com}[1]{}
\newcommand{\comg}[1]{}
\newcommand{\response}[1]{}
\fi

\title{\huge {Intelligent Reflecting Surface-Aided LEO Satellite Communication: Cooperative Passive Beamforming and Distributed Channel Estimation}}

\author{ 
	Beixiong Zheng,~\IEEEmembership{Member,~IEEE},
	Shaoe Lin, 
	and Rui Zhang,~\IEEEmembership{Fellow,~IEEE} 
	
	\thanks{\emph{(Corresponding author: Rui Zhang.)}
		
		B. Zheng and R. Zhang are with the Department of Electrical and Computer Engineering, National University of Singapore, Singapore 117583 (e-mail: \{elezbe, elezhang\}@nus.edu.sg).

	S. Lin is  with the School of Electronic and Information Engineering, South China University of Technology, Guangzhou
	510641, China (e-mail: eeshe.lin@mail.scut.edu.cn)
	}
}

\begin{document}
\maketitle
\begin{abstract}
Low-earth orbit (LEO) satellite communication plays an important role in assisting/complementing terrestrial communications by providing worldwide coverage, especially in harsh environments such as high seas, mountains, and deserts which are uncovered by terrestrial networks.
Traditionally, the passive reflect-array with fixed phase shifts has been applied in satellite communications to compensate for the high path loss due to long propagation distance with low-cost directional beamforming; however, it is unable to flexibly adapt the beamforming direction to dynamic channel conditions.
In view of this, we consider in this paper a new intelligent reflecting surface (IRS)-aided LEO satellite communication system, by utilizing the controllable phase shifts of massive passive reflecting elements to achieve flexible beamforming, which copes with the time-varying channel between the high-mobility satellite (SAT) and ground node (GN) cost-effectively.
In particular, we propose a new architecture for IRS-aided LEO satellite communication where IRSs are deployed at both sides of the SAT and GN, and study their cooperative passive beamforming (CPB) design over line-of-sight (LoS)-dominant single-reflection and double-reflection channels. Specifically,
we jointly optimize the active transmit/receive beamforming at the SAT/GN as well as the CPB at two-sided IRSs to maximize the overall channel gain from the SAT to each GN.
Interestingly, we show that under LoS channel conditions, the high-dimensional SAT-GN channel can be decomposed into the outer product of two low-dimensional vectors. By exploiting the decomposed SAT-GN channel, we decouple the original beamforming optimization problem into two simpler subproblems corresponding to the SAT and GN sides, respectively, which are both solved in closed-form.
Furthermore, we propose an efficient transmission protocol to conduct channel estimation and beam tracking, which only requires independent processing of the SAT and GN in a distributed manner, thus substantially reducing the implementation complexity.
 Simulation results validate the performance advantages of the proposed IRS-aided LEO satellite communication system with two-sided cooperative IRSs, as compared to various baseline schemes such as the conventional reflect-array and one-sided IRS.
\end{abstract}
\IEEEpeerreviewmaketitle

\section{Introduction}
To date, the fifth-generation (5G) wireless communication network has been rolled out in many countries and started changing people's lives dramatically. Meanwhile, the outbreak of the COVID-19 pandemic has made the public aware of the indispensable role of information and communications technology (ICT) in keeping our society running and connected.
Driven by the rapid growth of Internet-of-things (IoT) applications,
5G terrestrial communication network has been significantly advanced to meet the key performance requirements for enhanced mobile broadband (eMBB), ultra-reliable and low-latency communications (URLLC), and massive
machine-type communications (mMTC) \cite{andrews2014what,boccardi2014five,shafi20175g}. 
However, for remote regions such as deserts, mountains, rural areas, and oceans, the communication coverage by today's terrestrial networks is still widely unavailable due to the high difficulty and/or cost of deploying terrestrial base stations (BSs) and backhauls \cite{kapovits2018satellite,li2020enabling,Fang20215G}.
In view of this limitation, satellite communications have become a promising solution to assist/complement terrestrial communications, aiming to provide global coverage to support ubiquitous and seamless communications \cite{Fang20215G,Giordani2021Non,Chen2020Vision}.
Compared to terrestrial communication, satellite communication
achieves much wider coverage at lower cost and higher flexibility. 
Moreover, satellite is able to provide a variety of data services such as
unicast, multicast, broadcast, and relaying based on the requirements of ground nodes (GNs) on the Earth.
Therefore, it is anticipated that the upcoming sixth-generation (6G) wireless network will establish a fully connected world by integrating both terrestrial and satellite communications to take advantages of both \cite{Fang20215G,Giordani2021Non,Chen2020Vision}. 


Among various types of satellites operating at different orbital altitudes, the low-earth orbit (LEO) satellite has received the most attention recently for providing communication services to various GNs, due to its low altitude and the resultant advantages such as less  propagation loss, shorter transmission delay, and lower cost as compared to other higher orbit satellites  \cite{Fang20215G,Chen2020Vision}.
Nevertheless, \rev{the communication distance of the LEO satellite is still much larger than that of the terrestrial communication, thus suffering from orders-of-magnitude higher path loss.} As such, high-power transmitter and high-sensitive receiver \cite{Fang20215G} are usually applied in the LEO satellite communication to compensate for the high path loss, which, however, inevitably increases the hardware cost as well as power consumption. 
Alternatively, 
traditional passive reflect-array \cite{Mousavi2008LowCost,ma2021development,imaz2020reflectarray,hum2013reconfigurable} has been applied in the satellite/deep-space communication to achieve directional transmission for path loss compensation, without significantly increasing the hardware cost and power consumption. \rev{However, passive reflect-array provides fixed phase shifts once fabricated or a limited number of {\it coarse-grained}
passive beam patterns, which cannot be adaptively or flexibly changed when the LEO satellite flies over a target service region serving many GNs located at different locations and thus can incur performance degradation due to beam misalignment.}

Recently, the technological advances in micro electromechanical systems (MEMS) and metasurfaces \cite{cui2014coding,liaskos2018new,liu2018programmable} have made it feasible to reconfigure the passive reflecting surface in real time to dynamically adjust the passive signal reflection for creating favorable wireless propagation channels in a cost-effective manner, which leads to a promising new technology termed as intelligent reflecting surface (IRS) (or its various equivalents) \cite{wu2021intelligent,qingqing2019towards,Renzo2019Smart,basar2019wireless,zheng2021survey}.
Generally speaking, IRS is a large electromagnetic metasurface consisting of massive low-cost passive reflecting elements, each of which is able to independently control the reflection phase shift and/or amplitude of the incident signal so as to \rev{collaboratively enhance the reflected signal power significantly in desired directions (i.e., offering fine-grained passive beamforming gains).}
Owing to its reconfigurability in real time, IRS provides a new solution to cope with the time-varying wireless channel induced by the high-mobility LEO satellite as well as the severe path loss resulting from its long propagation distance.
Moreover, different from traditional active beamforming for satellite communications, IRS only uses tunable/controllable passive reflecting elements to achieve full-duplex passive beamforming at low hardware and energy cost, without the need of installing active antennas/radio-frequency (RF) chains for signal processing/amplification.

The appealing features of IRS have motivated a great deal of research on its application to various communication systems, e.g., orthogonal frequency division multiplexing (OFDM) \cite{zheng2019intelligent,zheng2020intelligent,zheng2020fast,yang2019intelligent}, relaying communication \cite{zheng2021irs,Yildirim2021Hybrid,nguyen2021hybrid}, and non-orthogonal multiple access (NOMA) \cite{Zheng2020IRSNOMA,yanggang2019intelligent,mu2021joint,zeng2020sum}, among others. Recently, some prior works have introduced the basic concept of IRS to satellite communication, where the IRS is placed either on the satellite \cite{tekbiyik2020reconfigurable,tekbiyik2020reconfigurable2} or near the ground users \cite{Dong2021Towards,xu2021intelligent,qian2021multi} to enhance their
communication performance. However, these works assumed the perfect channel state information (CSI) for passive beamforming design and performance analysis, but did not consider the important channel estimation/beam tracking issue for IRS, which is a challenging task in practice due to its large number of reflecting elements that do not possess signal transmission/processing capabilities, especially for the case of LEO satellite with high mobility.
\rev{Although practical passive beamforming designs and efficient channel estimation schemes (see, e.g., \cite{zheng2019intelligent,zheng2020intelligent,zheng2020fast,you2019progressive,Huang2019Reconfigurable,Wu2019TWC,yang2020energy,Pan2020Multicell,gong2021unified}) have been developed for various IRS-aided systems,
all of them are for the terrestrial communication with BSs and IRSs deployed at fixed locations to serve static/low-mobility users only, which may not be applicable to the high-mobility LEO satellite communication.}
For example, \rev{the high mobility of an LEO satellite induces the highly time-varying wireless channels with GNs and
can result in frequent channel estimation for active/passive beam alignment with them}, thus severely degrading the communication performance.
Furthermore, due to the large Doppler effect induced by the LEO satellite's high mobility, the channel estimation and beam tracking problems become more challenging in the IRS-aided LEO satellite communication.

\rev{Furthermore, most of the prior works on IRS have considered the wireless system aided by one or more non-cooperative IRSs (i.e., each independently serving its associated users) \cite{Wu2019TWC,Huang2019Reconfigurable,yang2020energy,Pan2020Multicell,gong2021unified,Zhang2020Capacity,Gao2020Distributed}; while the inter-IRS channel was usually ignored for simplicity.
Only recently, the great potential of the cooperative passive beamforming (CPB) over inter-IRS channels has been unveiled in the double-/multi-IRS aided communication system \cite{Han2020Cooperative,Zheng2020DoubleIRS,mei2021multi,you2020deploy,Zheng2021Efficient,zheng2020uplink,han2021double}, which was shown to achieve a much higher-order passive beamforming gain than its single-IRS counterpart.} In \cite{Han2020Cooperative}, the authors presented an initial study to reveal the CPB gain in a double-IRS aided communication system, where a single-antenna BS communicates with a single-antenna user over the double-reflection link established by two distributed IRSs placed near the BS and user, respectively.
By assuming the line-of-sight (LoS) channel model for the inter-IRS link, an $M^4$-fold CPB gain is achieved over the double-reflection link with a total number of $M$ passive reflecting elements, which significantly outperforms the conventional single-IRS baseline system with an $M^2$-fold passive beamforming gain over the single-reflection link only.
However, as compared to the single-reflection link, the double-reflection link generally suffers from higher product-distance path loss. As such, a sufficiently large $M$ is needed for the double-IRS system to beat its single-IRS counterpart with the same $M$, so that the high CPB gain can overcome the severe path loss in the double-reflection link to yield a higher channel gain.
Later, the authors of \cite{Zheng2020DoubleIRS} considered the more general double-IRS aided communication system with a multi-antenna BS in the presence of both double- and single-reflection links. In contrast to \cite{Han2020Cooperative} that focused on the double-reflection link only, the work in \cite{Zheng2020DoubleIRS} made use of both double- and single-reflection links to achieve their coherent channel combination, thus further improving the performance of the double-IRS aided system against the single-IRS counterpart.

 \begin{figure}[!t]
	\centering
	\includegraphics[width=6in]{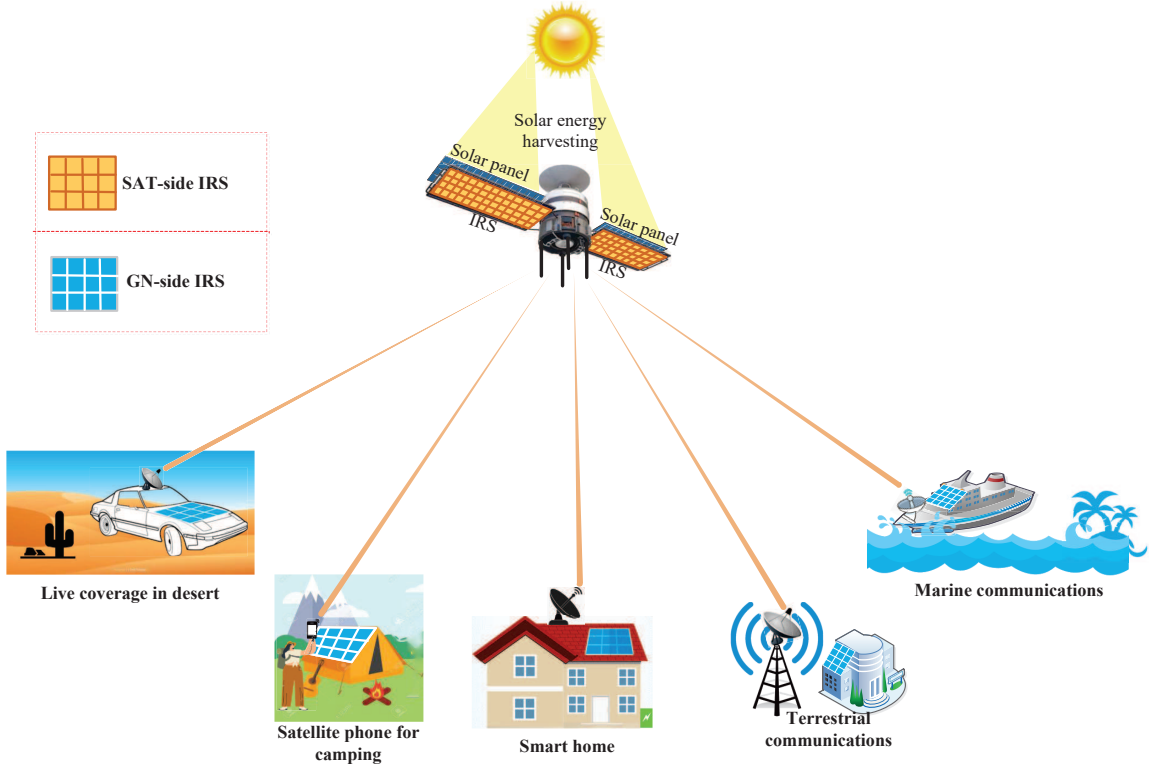}
	\setlength{\abovecaptionskip}{-3pt}
	\caption{IRS-aided LEO satellite communications with both SAT-side and GN-side IRSs.}
	\label{systemAPP}
\end{figure}
Motivated by the above, we consider in this paper a new IRS-aided LEO satellite communication system with two-sided cooperative IRSs as shown in Fig.~\ref{systemAPP}, where the communications between the LEO satellite (SAT) and various GNs in different applications/scenarios are aided by distributed IRSs deployed near them, thus referred to as the SAT-side IRS and GN-side IRS, respectively.
Specifically, the LEO satellite is equipped with solar panels for harvesting solar energy from the Sun, while 
the SAT-side IRSs are coated on the reverse side of these solar panels and thus face towards the Earth for assisting the satellite communication. On the other hand, depending on the specific communication scenario on the ground, the GN-side IRSs can be flexibly mounted on cars, camping tents, house roofs, buildings, ships, etc., as shown in Fig.~\ref{systemAPP}.
It is noted that \rev{due to the high altitude of the SAT, its channels to different GNs as well as  their nearby IRSs are very likely to be LoS in practice \cite{Wang20206G}, which helps achieve more pronounced passive beamforming gains over both the single-reflection and double-reflection links via two-sided IRSs between the SAT and GNs.
Besides, different from other high-mobility communications (such as the UAV communication), the orbit of the LEO satellite
is highly predictable, which can be exploited to achieve efficient beam tracking.}
Assuming the LoS channel model for all the involved links (including the direct, single-reflection, and double-reflection links), we jointly optimize the active transmit/receive beamforming at the SAT/GN and the CPB at two-sided IRSs to maximize the overall channel gain from the SAT to each GN for data transmission.
Moreover, as the LEO satellite operates in high mobility, active/passive beam alignment in real time is of paramount importance to maintain a high overall channel gain for the joint active and passive beamforming.
\rev{Although there are some existing works (see, e.g., \cite{arapoglou2010mimo,zhao2018beam,jeon2000new}) on beam tracking for satellites, they have been proposed for active transceivers only and thus are inapplicable to our considered satellite system that needs to track both active and passive beams at the transceivers and IRSs.
In particular, due to a large number of reflecting elements without signal transmission/processing capabilities, the passive beam tracking at IRSs is much more challenging.
As such, it calls for more efficient channel estimation and beam tracking methods tailored for the high-mobility satellite communication with two-sided IRSs.}
The main contributions of this paper are summarized as follows.
\begin{itemize}
	\item First, to efficiently solve the joint optimization problem for the active and passive beamforming in the considered IRS-aided LEO satellite communication, \rev{we show that
	the high-dimensional SAT-GN multiple-input-multiple-output (MIMO) channel matrix can be decomposed into the outer product of two low-dimensional vectors under LoS channel conditions.
	By exploiting the decomposed SAT-GN channel, we decouple the original beamforming optimization problem into two simpler subproblems corresponding to the SAT and GN sides, respectively, which are both optimally solved in closed-form with very low complexity.}
	\item Second, based on the channel decomposition result, we show that the local CSI at the SAT and GN sides is sufficient for solving the two decoupled beamforming optimization problems, respectively.
	Motivated by this, we propose an efficient transmission protocol to estimate the local CSI separately at the SAT and GN sides based on the uplink and downlink training, respectively, \rev{which requires their independent processing only and achieves very low training overhead.}
	In addition, to avoid the high-complexity maximum likelihood (ML) estimation, \rev{we further propose a least-square (LS) based channel estimation scheme to estimate the required local CSI at the SAT/GN side in a decoupled manner, thus significantly reducing the complexity for practical implementation.}
	\item Third, after initial channel estimation, we also propose an efficient distributed beam tracking scheme to lock on the time-varying channels due to the LEO satellite's high mobility. Specifically, by exploiting the predictable orbit of the LEO satellite with the {\it prior} information on its altitude and velocity, the active and passive beam directions at both the SAT and GN sides can be efficiently updated over time in a distributed manner by assuming a linear time-varying model, thus achieving accurate beam alignment in real time to maintain a high overall channel gain for data transmission.
	\item Last, we provide extensive numerical results to validate the performance superiority of our proposed LEO satellite communication aided with two-sided cooperative IRSs to various baseline schemes such as the conventional reflect-array and one-sided IRS. Besides, the effectiveness of our proposed distributed channel estimation and beam tracking schemes is also verified by simulation results. 
\end{itemize}

The rest of this paper is organized as follows. Section~\ref{sys} presents the system model for the IRS-aided LEO satellite communication with two-sided cooperative IRSs. In Section~\ref{design}, we propose an efficient design to solve the joint active and passive beamforming optimization problem for the considered system. 
In Section~\ref{CH_Est}, we propose a practical transmission protocol to conduct distributed channel estimation and beam tracking at the SAT and GN sides efficiently. 
Simulation results are presented in Section~\ref{Sim} to evaluate the performance of the
proposed system and practical designs. Finally, conclusions are drawn in Section~\ref{conlusion}.

\emph{Notation:} 
Upper-case and lower-case boldface letters denote matrices and column vectors, respectively.
Upper-case calligraphic letters (e.g., $\cal{F}$) denote discrete and finite sets.
Superscripts ${\left(\cdot\right)}^{T}$, ${\left(\cdot\right)}^{H}$, ${\left(\cdot\right)}^{*}$, ${\left(\cdot\right)}^{-1}$, and ${\left(\cdot\right)}^{\dagger}$ stand for the transpose, Hermitian transpose, conjugate, matrix inversion, and pseudo-inverse operations, respectively.
${\mathbb C}^{a\times b}$ denotes the space of ${a\times b}$ complex-valued matrices.
For a complex-valued vector $\bm{x}$, $\lVert\bm{x}\rVert$ denotes its $\ell_2 $-norm,
$\angle (\bm{x} )$ returns the phase of each element in $\bm{x}$,
and ${\rm diag} (\bm{x})$ returns a diagonal matrix with the elements in $\bm{x}$ on its main diagonal.
$|\cdot|$ denotes the absolute value if applied to a complex-valued number or the cardinality if applied to a set.
${\cal O}(\cdot)$ stands for the standard big-O notation,
$\otimes$ denotes the Kronecker product, 
and $\odot$ denotes the Hadamard product.
$[{\bm A}]_{i,j}$ denotes the $(i,j)$-th entry of matrix ${\bm A}$, while
${\bm I}$ and ${\bm 0}$ denote an identity matrix and an all-zero matrix, respectively, with appropriate dimensions.
The distribution of a circularly symmetric complex Gaussian (CSCG) random vector with zero-mean and covariance matrix ${\bm \Sigma}$ is denoted by ${\mathcal N_c }({\bm 0}, {\bm \Sigma} )$; and $\sim$ stands for ``distributed as".

\section{System Model}\label{sys}
\begin{figure}[!t]
 	\centering
 	\includegraphics[width=5.5in]{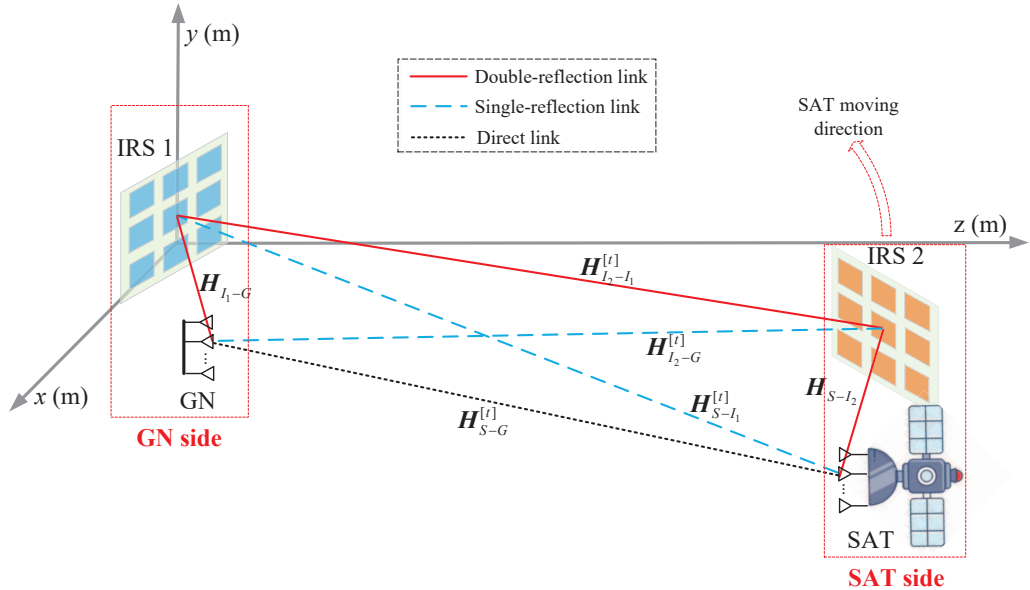}
 	\setlength{\abovecaptionskip}{-3pt}
 	\caption{An illustration of the LEO satellite communication with two-sided cooperative IRSs (3D view).}
 	\label{system}
 \end{figure}

As shown in Fig. \ref{system}, we consider an IRS-aided LEO satellite communication with two-sided cooperative IRSs, in which the communication between a GN\footnote{For the purpose of exposition, we consider the satellite communication with one served GN where no co-channel interference with other GNs is present, which corresponds to the practical scenario when orthogonal multiple access (such as time division multiple access (TDMA)) is employed to separate the communications for different GNs.}
 and an LEO satellite is aided by two IRSs deployed on/near each of them, respectively.
 We assume that the GN and SAT are equipped with uniform planar arrays (UPAs) that consist of $N_{1}\triangleq N_{1,x}\times N_{1,y}$ and $N_{2}\triangleq N_{2,x}\times N_{2,y}$ active antennas, respectively.
 \rev{Moreover, the GN-side IRS is connected to a smart controller via a wire link and deployed near the GN,} which can be regarded as one equivalent (passive) UPA equipped with $M_{1}\triangleq M_{1,x}\times M_{1,y}$ passive reflecting elements and is labeled as IRS 1. 
 On the other hand, the SAT-side IRS is mounted on the reverse side of the satellite solar panels, which can be modeled as one equivalent (passive) UPA equipped with totally $M_{2}\triangleq M_{2,x}\times M_{2,y}$ passive reflecting elements and is labeled as IRS~2. \rev{In particular, the SAT-side IRS can be integrated with the transceiver at the satellite via a wire link. As such, the transceiver at the satellite takes over the role of the conventional IRS controller for adjusting the SAT-side IRS
 reflection in real time.
Furthermore, under the three-dimensional (3D) Cartesian coordinate system shown in Fig. \ref{system}, we assume that the LEO satellite with IRS~2 orbits the Earth (assumed to be an ideal sphere) at a fixed altitude of $L_S$. Let $L_O$ denote the radius of the satellite's orbit and we have $L_O=L_S+L_E$ with $L_E$ being the Earth's radius.}

\begin{figure}[!t]
	\centering
	\includegraphics[width=5.5in]{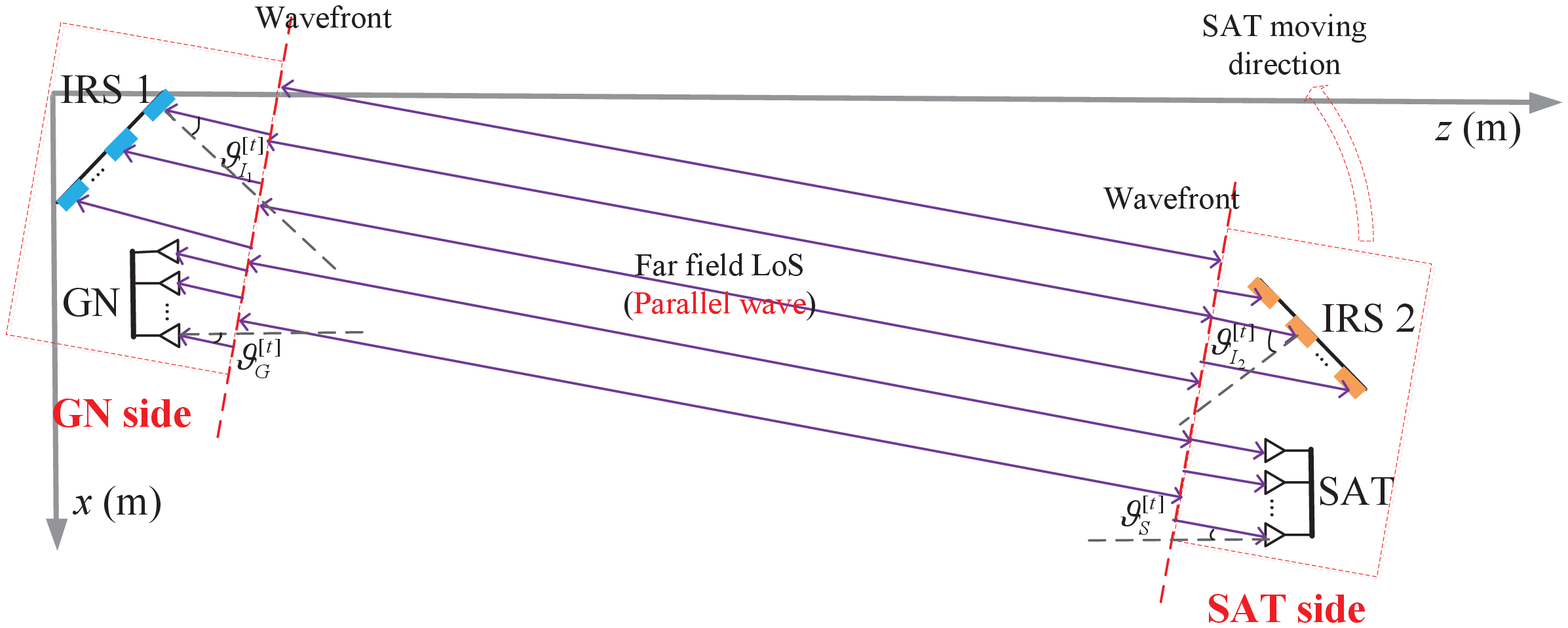}
	\setlength{\abovecaptionskip}{-3pt}
	\caption{An illustration of the wave propagation and AoAs/AoDs between the GN and SAT sides (2D view).}
	\label{system2D}
\end{figure}

 As shown in Fig. \ref{system}, due to the relative motion between the GN and SAT (dominated by the high-speed satellite), we denote  ${{\bm H}}_{S-G}^{[t]} \in {\mathbb{C}^{N_1\times N_2 }}$, ${{\bm H}}_{S-I_1}^{[t]} \in {\mathbb{C}^{M_1\times N_2 }}$, ${{\bm H}}_{I_2-G}^{[t]} \in {\mathbb{C}^{N_1\times M_2 }}$, and ${{\bm H}}_{I_2-I_1}^{[t]}  \in {\mathbb{C}^{M_1\times M_2 }}$
 as the baseband equivalent time-varying channels for the SAT$\rightarrow$GN, SAT$\rightarrow$IRS~1, IRS~2$\rightarrow$GN, and IRS~2$\rightarrow$IRS~1 links at time $t$, respectively.\footnote{For notational convenience, we use subscripts ''$G$", ''$I_1$", ''$S$", and ''$I_2$" to indicate the GN, IRS~1, SAT, and IRS~2, respectively.}
 Due to the high altitude of the SAT side, the channels from the SAT and IRS~2 to the GN and IRS 1, i.e., $\left\{ {{\bm H}}_{S-G}^{[t]}, {{\bm H}}_{S-I_1}^{[t]}, {{\bm H}}_{I_2-G}^{[t]}, {{\bm H}}_{I_2-I_1}^{[t]} \right\}$ are assumed to be LoS \cite{Wang20206G}.
Accordingly, these MIMO channels are all of rank-one.
For convenience, we first define a
one-dimensional (1D) steering vector function for a generic uniform linear array (ULA) as follow.
 \begin{align}
{\bm e}(\phi,N)\triangleq\left[1, e^{-j \pi \phi},\ldots, e^{-j \pi(N-1) \phi}\right]^T\in {\mathbb{C}^{N \times 1 }},
\end{align}
where $j\triangleq \sqrt{1}$ denotes the imaginary unit,
$\phi$ denotes the constant phase-shift difference between the signals at two adjacent antennas/elements, and
$N$ denotes the number of antennas/elements in the ULA. 
Note that due to the extremely long distance between the GN and SAT sides\footnote{An LEO satellite typically orbits the earth at altitude ranging from about $500$ to $2,000$ km.}, the propagation channels between the two sides can be characterized by the far-field LoS model with {\it parallel wavefronts}, as illustrated in Fig.~\ref{system2D}.
Accordingly, we let ${\bm a}_G( \vartheta_G^{[t]}, \varphi_G^{[t]})$, ${\bm a}_S( \vartheta_S^{[t]}, \varphi_S^{[t]})$, ${\bm a}_{I_1}( \vartheta_{I_1}^{[t]}, \varphi_{I_1}^{[t]})$,
and ${\bm a}_{I_2}( \vartheta_{I_2}^{[t]}, \varphi_{I_2}^{[t]})$ denote the array response vectors 
associated with the angle-of-arrival/departure (AoA/AoD) pairs $( \vartheta_G^{[t]}, \varphi_G^{[t]})$, $( \vartheta_S^{[t]}, \varphi_S^{[t]})$, $( \vartheta_{I_1}^{[t]}, \varphi_{I_1}^{[t]})$, and $( \vartheta_{I_2}^{[t]}, \varphi_{I_2}^{[t]})$
of the GN, SAT, IRS~1, and IRS~2 at time $t$, respectively.
Under the UPA model, each array response vector is expressed as the Kronecker product of two steering vector functions in the $x$-axis (horizontal) and $y$-axis (vertical) directions, respectively. For example, the array response vector at the GN is expressed as
 \begin{align}
\hspace{-0.2cm}{\bm a}_G( \vartheta_G^{[t]}, \varphi_G^{[t]})={\bm e}\left( \hspace{-0.1cm}\frac{2\Delta_G}{\lambda} \cos (\varphi_G^{[t]})\cos (\vartheta_G^{[t]})  , N_{1,x} \hspace{-0.1cm}\right)\hspace{-0.1cm} \otimes\hspace{-0.1cm} {\bm e}\left(\hspace{-0.1cm}\frac{2\Delta_G}{\lambda} \cos(\varphi_G^{[t]})\sin (\vartheta_G^{[t]}) , N_{1,y} \hspace{-0.1cm}\right)\hspace{-0.1cm}\in\hspace{-0.1cm} {\mathbb{C}^{N_1 \times 1 }},
\end{align}
with $\lambda$ denoting the signal wavelength and
$\Delta_G$ being the antenna spacing at the GN; while the other array response vectors can be similarly defined.
Accordingly, the real-time far-field LoS channels between any two nodes (represented by $X$ and $Y$ for notational simplicity) are modeled as the outer product of array responses at their two sides, i.e., 
\begin{align}\label{Far_LoS}
{{\bm H}}_{Y-X}^{[t]}&=\underbrace{\frac{\sqrt{\beta}}{d_{X-Y}^{[t]}} e^{\frac{-j 2 \pi  }{\lambda}d_{X-Y}^{[t]}}}_{\rho_{X-Y}^{[t]} }   {\bm a}_X( \vartheta_X^{[t]}, \varphi_X^{[t]})  {\bm a}^T_Y( \vartheta_Y^{[t]}, \varphi_Y^{[t]}),\qquad  X \in \{G, I_1\},~~ Y \in \{S, I_2\},
\end{align}
where $\beta$ stands for the reference path gain at the distance of 1 meter (m),
$d_{X-Y}^{[t]} $ denotes the real-time propagation distance between nodes $X$ and $Y$, and $\rho_{X-Y}^{[t]}$ is the corresponding complex-valued path gain between them at time $t$.

We then let ${{\bm H}}_{I_1-G}\in {\mathbb{C}^{N_1\times M_1}}$ and ${{\bm H}}_{S-I_2}\in {\mathbb{C}^{M_2\times N_2}}$ denote the baseband equivalent channels for the IRS~1$\rightarrow$GN and SAT$\rightarrow$IRS~2 links at the GN and SAT sides, respectively.
It is worth pointing out that the GN and IRS~1 are assumed to be at given locations with a fixed distance.
On the other hand, despite their high mobility, the SAT and IRS~2 remain relatively static with a fixed distance.
As such, due to the above fixed geometric relationship and the limited scattering environment, we model the IRS~1$\rightarrow$GN and SAT$\rightarrow$IRS~2 links, i.e., $\left\{ {{\bm H}}_{I_1-G}, {{\bm H}}_{S-I_2}\right\}$ as time-invariant LoS channels that are determined by their fixed distances.
For example, the $(n, m)$-th channel coefficient of the IRS~1$\rightarrow$GN channel ${{\bm H}}_{I_1-G}$ is given by
\begin{align}\label{GN_IRS1}
	\left[{{\bm H}}_{I_1-G} \right]_{n, m}=\frac{\sqrt{\beta}}{d_{n, m}} e^{\frac{-j 2 \pi }{\lambda} d_{n, m}},\qquad  n=1, \ldots N_1, \quad m=1, \ldots M_1, 
\end{align}
where $d_{n, m}$ is the distance between the $n$-th antenna of the GN and the $m$-th reflecting element of IRS~1. Following the above, the channel coefficients of the SAT$\rightarrow$IRS~2 channel ${{\bm H}}_{S-I_2}$ can be similarly defined.
It is noted that we assume all the above links on/between the GN and SAT sides are LoS regardless of whether they are time-variant or not; as a result, channel reciprocity holds for each link in the uplink and downlink communications.  The effect of non-LoS (NLoS) channel components on the system performance will be evaluated in Section~\ref{Sim}. 

Let ${\bm \theta}_\mu^{[t]}\triangleq[{\theta_{\mu,1}^{[t]}},{\theta_{\mu,2}^{[t]}},\ldots,{\theta_{\mu,M_\mu}^{[t]}}]^T$ denote the equivalent tunable reflection coefficients of IRS $\mu$ at time $t$, where the reflection amplitudes of all reflecting elements are set to one or the maximum value to maximize the signal reflection power as well as ease the hardware implementation, leading to $\left|{\theta_{\mu,m}^{[t]}}\right|=1, \forall m=1,\ldots,M_\mu$, $\mu\in \{1,2\}$.
Under the above setup, the overall/effective channel matrix from the SAT to the GN at time $t$ is given by
\begin{align}\label{Eff_Ch}
{\bar{\bm H}}^{[t]}=\underbrace{ {{\bm H}}_{S-G}^{[t]}}_{\rm Direct~link}+ 
\underbrace{ {{\bm H}}_{I_1-G} {\bm \Theta}_1^{[t]} {{\bm H}}_{S-I_1}^{[t]} + {{\bm H}}_{I_2-G}^{[t]} {\bm \Theta}_2^{[t]} {{\bm H}}_{S-I_2}  }_{\rm Single-reflection~links}+
\underbrace{ {{\bm H}}_{I_1-G}{\bm \Theta}_1^{[t]} {{\bm H}}_{I_2-I_1}^{[t]} {\bm \Theta}_2^{[t]} {{\bm H}}_{S-I_2} }_{\rm Double-reflection~link},
\end{align}
where ${\bm \Theta}_\mu^{[t]}={\rm diag} \left( {\bm \theta}_\mu^{[t]} \right)$ represents the diagonal reflection matrix of IRS $\mu$, $\mu\in \{1,2\}$, at time $t$.
According to \eqref{Eff_Ch}, in order to jointly design the CPB vectors $\left\{{\bm \theta}_1^{[t]}, {\bm \theta}_2^{[t]}\right\}$ at the two-sided IRSs for enhancing the uplink/downlink data transmission in the IRS-aided satellite communication, we need to acquire the real-time CSI of $\left\{{{{\bm H}}_{S-G}^{[t]}}, {{\bm H}}_{S-I_1}^{[t]}, {{\bm H}}_{I_2-G}^{[t]}, {{\bm H}}_{I_2-I_1}^{[t]}, {{\bm H}}_{I_1-G}, {{\bm H}}_{S-I_2} \right\}$.
Among them, the time-invariant IRS~1$\rightarrow$GN and SAT$\rightarrow$IRS~2 channels $\left\{ {{\bm H}}_{I_1-G}, {{\bm H}}_{S-I_2}  \right\}$ can be obtained offline (or updated over a long period) according to \eqref{GN_IRS1} based on the fixed LoS distances, which are thus assumed to be known at the GN and SAT, respectively.
However, due to the rapid relative motion between the GN and SAT sides,
the other channels $\left\{{{{\bm H}}_{S-G}^{[t]}}, {{\bm H}}_{S-I_1}^{[t]}, {{\bm H}}_{I_2-G}^{[t]}, {{\bm H}}_{I_2-I_1}^{[t]}\right\}$ are highly dynamic, which need to be efficiently estimated and tracked in real time for designing the active and passive beamforming, in order to maximize the effective channel gain in \eqref{Eff_Ch} over time. 

In the following, \rev{we first focus on characterizing the optimal performance gain brought by the IRS-aided satellite communication with two-sided cooperative IRSs in Section~\ref{design} by jointly optimizing the active and passive beamforming under the assumption of perfect CSI, which thus serves as the performance upper bound.
Then, to tackle the CSI acquisition/tracking issue in practice, we propose a new distributed channel estimation scheme with low training overhead as well as an effective beam tracking scheme tailored for the considered high-mobility satellite communication with two-sided IRSs in Section~\ref{CH_Est}.}

\section{Joint Active and Passive Beamforming Design}\label{design}
In this section, we aim at deriving the optimal active and passive beamforming design to maximize the effective SAT-GN channel gain and \rev{draw essential insights under the assumption that the perfect real-time CSI of all the relevant links is available.}
For brevity, we drop the time index ${[t]}$ in this section without causing any confusion.
Let ${\bm w}_1\in{\mathbb{C}^{N_1 \times 1 }}$ and ${\bm w}_2\in{\mathbb{C}^{N_2 \times 1 }}$ denote the active transmit/receive beamforming vectors at the GN and SAT, respectively, and apply them to the channel model in \eqref{Eff_Ch}, which yields the effective SAT-GN channel gain as
\begin{align}\label{CH_gain}
\gamma\left({\bm \theta}_1,{\bm w}_1, {\bm \theta}_2,{\bm w}_2\right) =\left| {\bm w}_1^T{\bar{\bm H}}  {\bm w}_2\right|^2,
\end{align}
where we have $\|{\bm w}_1\|^2=1$ and $\|{\bm w}_2\|^2=1$ for normalization. Accounting for the constraints on the active and passive beamforming, the optimization problem for maximizing the overall channel gain in \eqref{CH_gain} is formulated as follows.
\begin{align}
\text{(P1):}~
& \underset{ {\bm \theta}_1,{\bm w}_1, {\bm \theta}_2,{\bm w}_2 }{\text{max}}
& & \gamma\left({\bm \theta}_1,{\bm w}_1, {\bm \theta}_2,{\bm w}_2\right)\label{obj_P0}\\
& \text{~~~~s.t.} & & |{\theta_{\mu,m}}|=1, \quad~ \forall m=1,\ldots,M_\mu, \mu\in \{1,2\}, \qquad\qquad\qquad \label{con1_P0}\\
& & & \|{\bm w}_\mu\|^2=1, \quad \forall \mu\in \{1,2\} \label{con2_P0}.
\end{align}
It can be verified that problem (P1) is a non-convex optimization problem due to the unit-modulus
constraint in \eqref{con1_P0} and the fact that the active transmit/receive beamforming vectors (i.e., ${\bm w}_1$ and ${\bm w}_2$) at the GN and SAT are coupled with the passive beamforming vectors (i.e., ${\bm \theta}_1$ and ${\bm \theta}_2$) at the two distributed IRSs in the objective function of \eqref{obj_P0}. Although this non-convex optimization problem is generally difficult to be solved, we first simplify it by exploiting the unique channel structure of the IRS-aided satellite communication with two-sided IRSs. 

To this end, by substituting \eqref{Far_LoS} into \eqref{Eff_Ch}, the channel model in \eqref{Eff_Ch} can be expressed as
\begin{align}\label{Eff_Ch2}
&{\bar{\bm H}}=\rho_{G-S} {\bm a}_G( \vartheta_G, \varphi_G)  {\bm a}^T_S( \vartheta_S, \varphi_S)+ 
\rho_{I_1-S} {{\bm H}}_{I_1-G} {\bm \Theta}_1  {\bm a}_{I_1}( \vartheta_{I_1}, \varphi_{I_1})  {\bm a}^T_S( \vartheta_S, \varphi_S)  +  \rho_{G-I_2}\times\notag\\
 &{\bm a}_G( \vartheta_G, \varphi_G)  {\bm a}^T_{I_2}( \vartheta_{I_2}, \varphi_{I_2}) {\bm \Theta}_2 {{\bm H}}_{S-I_2}
  +\rho_{I_1-I_2} {{\bm H}}_{I_1-G} {\bm \Theta}_1  {\bm a}_{I_1}( \vartheta_{I_1}, \varphi_{I_1})  {\bm a}^T_{I_2}( \vartheta_{I_2}, \varphi_{I_2}) {\bm \Theta}_2 {{\bm H}}_{S-I_2}.
\end{align}
Then we decompose the complex-valued path gain $\rho_{X-Y}$ as $\rho_{X-Y}\triangleq \rho_{X}\cdot\rho_{Y}$ with $X \in \{G, I_1\}, Y \in \{S, I_2\}$, and the channel model in \eqref{Eff_Ch2} can be equivalently expressed as
\begin{align}\label{Eff_Ch3}
\hspace{-0.1cm}{\bar{\bm H}}&=\Big(\rho_{G} {\bm a}_G( \vartheta_G, \varphi_G)+\rho_{I_1} {{\bm H}}_{I_1-G} {\bm \Theta}_1  {\bm a}_{I_1}( \vartheta_{I_1}, \varphi_{I_1})\Big)   \Big(\rho_{S} {\bm a}^T_S( \vartheta_S, \varphi_S) + \rho_{I_2} {\bm a}^T_{I_2}( \vartheta_{I_2}, \varphi_{I_2}) {\bm \Theta}_2 {{\bm H}}_{S-I_2} \Big)\notag\\
&=\hspace{-0.1cm}\underbrace{\left[
	\rho_{G}{\bf I}_{N_1}, \rho_{I_1} {{\bm H}}_{I_1-G} {\bm \Theta}_1 \right]\hspace{-0.15cm}
	\begin{bmatrix}
	\hspace{-0.2cm}{\bm a}_G( \vartheta_G, \varphi_G) \\
	{\bm a}_{I_1}( \vartheta_{I_1}, \varphi_{I_1})
	\hspace{-0.1cm}\end{bmatrix} }_{{\rm GN~side}:~\triangleq {\bm f}_1({\bm \theta}_1) }
\underbrace{\left[{\bm a}^T_S( \vartheta_S, \varphi_S), {\bm a}^T_{I_2}( \vartheta_{I_2}, \varphi_{I_2}) \right]
	\hspace{-0.2cm}\begin{bmatrix}
	\rho_{S}{\bf I}_{N_2} \\
	\hspace{-0.1cm}\rho_{I_2} {\bm \Theta}_2 {{\bm H}}_{S-I_2} \hspace{-0.1cm}
	\end{bmatrix}
}_{{\rm SAT~side}:~\triangleq {\bm f}^T_2({\bm \theta}_2)},\hspace{-0.3cm}
\end{align}
where ${\bar{\bm a}}_G\triangleq\left[{\bm a}^T_G( \vartheta_G, \varphi_G), {\bm a}^T_{I_1}( \vartheta_{I_1}, \varphi_{I_1})\right]^T\in {\mathbb{C}^{(N_1+M_1) \times 1 }}$ and 
${\bar{\bm a}}_S\triangleq\left[{\bm a}^T_S( \vartheta_S, \varphi_S), {\bm a}^T_{I_2}( \vartheta_{I_2}, \varphi_{I_2}) \right]^T \in {\mathbb{C}^{(N_2+M_2) \times 1 }}$ can be regarded as the joint active and passive array response vectors at the GN and SAT sides, respectively. 
It can be shown in \eqref{Eff_Ch3} that the effective SAT-GN channel matrix ${\bar{\bm H}}$ is decomposed into the outer product of two low-dimensional channel vectors $ {\bm f}_1({\bm \theta}_1)$ and ${\bm f}_2({\bm \theta}_2)$ at the GN and SAT sides, respectively. \rev{Accordingly, by substituting \eqref{Eff_Ch3}, i.e., ${\bar{\bm H}}={\bm f}_1({\bm \theta}_1){\bm f}^T_2({\bm \theta}_2)$ into \eqref{CH_gain},
the effective SAT-GN channel gain in \eqref{CH_gain} can be rewritten as
\begin{align}\label{decouple}
\gamma\left({\bm \theta}_1,{\bm w}_1, {\bm \theta}_2,{\bm w}_2\right) =\left| {\bm w}_1^T{\bm f}_1({\bm \theta}_1)   {\bm f}^T_2({\bm \theta}_2) {\bm w}_2\right|^2=\underbrace{\left| {\bm w}_1^T{\bm f}_1({\bm \theta}_1)\right|^2}_{\gamma_1\left({\bm \theta}_1,{\bm w}_1\right)} \cdot \underbrace{\left|  {\bm f}^T_2({\bm \theta}_2) {\bm w}_2\right|^2}_{\gamma_2\left({\bm \theta}_2,{\bm w}_2\right)}.
\end{align}}
From \eqref{decouple}, we see that the joint active and passive beamforming design for maximizing the overall channel gain $\gamma\left({\bm \theta}_1,{\bm w}_1, {\bm \theta}_2,{\bm w}_2\right)$ can be decoupled at the GN and SAT sides for maximizing $\gamma_1\left({\bm \theta}_1,{\bm w}_1\right)$ and $\gamma_2\left({\bm \theta}_2,{\bm w}_2\right)$, respectively, without loss of optimality.
Moreover, it can be readily verified that for any given passive beamforming vectors ${\bm \theta}_1$ and ${\bm \theta}_2$, the optimal active transmit/receive beamforming solution for maximizing $\gamma_1\left({\bm \theta}_1,{\bm w}_1\right)$ and $\gamma_2\left({\bm \theta}_2,{\bm w}_2\right)$ is  the maximum-ratio transmission/combination (MRT/MRC), i.e.,
\begin{align}\label{active}
{\bm w}^{\star}_\mu=\frac{{\bm f}^*_\mu({\bm \theta}_\mu)}{\|{\bm f}_\mu({\bm \theta}_\mu)\|}, \qquad \mu\in \{1,2\}.
\end{align}
As such, after substituting ${\bm w}^{\star}_\mu$ into $\gamma_\mu\left({\bm \theta}_\mu,{\bm w}_\mu\right)$,
the optimization problem in (P1) for the (remaining)  CPB design \rev{is equivalently transformed into the following subproblems:}
\begin{align}
\text{(P2):}~
\gamma^\star_\mu\triangleq& \underset{ {\bm \theta}_\mu}{\text{max}}
& & \left\| {\bm f}_\mu({\bm \theta}_\mu)\right\|^2\label{obj_P2}\\
& \text{s.t.} & & |{\theta_{\mu,m}}|=1, \quad~ \forall m=1,\ldots,M_\mu,\qquad\qquad\qquad\qquad \label{con1_P2}
\end{align}
with $\mu\in \{1,2\}$ corresponding to the passive beamforming design at the GN and SAT sides, respectively. As can be seen, each decoupled subproblem in (P2) is simpler than (P1) but still non-convex due to the unit-modulus constraint in \eqref{con1_P2}.
In the following, we focus on the passive beamforming design at the GN side with $\mu=1$ in problem (P2) for the purpose of exposition; while the solution holds for the SAT side with $\mu=2$ in problem (P2) similarly.

\rev{For the GN side with $\mu=1$, problem (P2) can be explicitly expressed in an equivalent form as
\begin{align}
\text{(P3):}~
\gamma^\star_1\triangleq& \underset{ {\bm \theta}_1}{\text{max}}
& & \Big\| \underbrace{\rho_{G} {\bm a}_G( \vartheta_G, \varphi_G)+\rho_{I_1} {{\bm H}}_{I_1-G} {\bm \Theta}_1  {\bm a}_{I_1}( \vartheta_{I_1}, \varphi_{I_1})}_{{\bm f}_1({\bm \theta}_1)} \Big\|^2\label{obj_P3}\\
& \text{s.t.} & & |{\theta_{1,m}}|=1, \quad~ \forall m=1,\ldots,M_1.\qquad\qquad\qquad\qquad\qquad \label{con1_P3}
\end{align}}
It turns out that problem (P3) has a similar formulation to that of the single-IRS aided multiple-input and single-output (MISO) system considered in \cite{Wu2019TWC}, which thus can be approximately solved by the semidefinite relaxation (SDR) algorithm proposed in \cite{Wu2019TWC}.
However, the SDR algorithm generally needs a large number of iterations to reach convergence and has a relatively high computational complexity 
in the order of ${\cal O} \left(M_1^{4.5} \right)$ for each iteration, which may not be implementable for the high-mobility satellite communication.
To tackle this issue, we propose a low-complexity passive beamforming design by exploiting the peculiar LoS channel structure of the considered system, elaborated as follows.

Similar to \cite{han2021double}, we reasonably consider that the link distance between the GN and IRS~1 is large enough (e.g., considerably larger than the array sizes at the GN and IRS~1) such that ${{\bm H}}_{I_1-G}$ in \eqref{GN_IRS1} can be well approximated as a rank-one LoS channel matrix given by
\begin{align}\label{rank-one}
{{\bm H}}_{I_1-G} \approxeq \underbrace{\frac{\sqrt{\beta}}{d_{I_1-G}} e^{\frac{-j 2 \pi }{\lambda} d_{I_1-G}}}_{\delta_{I_1-G}} {\bar{\bm h}}_{G} {\bar{\bm h}}^T_{I_1},
\end{align}
where $d_{I_1-G}$ accounts for the reference distance between the GN and IRS~1, $\delta_{I_1-G}$ is the corresponding complex-valued path gain, and
${\bar{\bm h}}_{G}\in {\mathbb{C}^{N_1 \times 1 }}$ and ${\bar{\bm h}}_{I_1}\in {\mathbb{C}^{M_1 \times 1 }}$ denote the two array responses at the GN and IRS~1, respectively.
\rev{By substituting \eqref{rank-one}, i.e., ${{\bm H}}_{I_1-G} \approxeq\delta_{I_1-G}{\bar{\bm h}}_{G} {\bar{\bm h}}^T_{I_1}$ into ${\bm f}_1({\bm \theta}_1)$ of \eqref{obj_P3}, we obtain}
\begin{align}\label{f1}
{\bm f}_1({\bm \theta}_1)&=\rho_{G} {\bm a}_G( \vartheta_G, \varphi_G)+\rho_{I_1} \delta_{I_1-G} {\bar{\bm h}}_{G} {\bar{\bm h}}^T_{I_1} {\bm \Theta}_1  {\bm a}_{I_1}( \vartheta_{I_1}, \varphi_{I_1})\notag\\
&=\rho_{G} {\bm a}_G( \vartheta_G, \varphi_G)+ c({\bm \theta}_1) {\bar{\bm h}}_{G},
\end{align}
where we define $c({\bm \theta}_1)\triangleq \rho_{I_1} \delta_{I_1-G} {\bar{\bm h}}^T_{I_1} {\bm \Theta}_1  {\bm a}_{I_1}( \vartheta_{I_1}, \varphi_{I_1})$ for notational convenience.
In particular, ${\bm \theta}_1$ should be set to maximize the passive beamforming gain in $\left|c({\bm \theta}_1)\right|$, so as to maximize the overall effective gain of $\left\|{\bm f}_1({\bm \theta}_1)\right\|^2$.
\rev{Accordingly, the optimal passive beamforming vector (denoted by ${\bm \theta}_1^{\star}$) at IRS~1 is designed as
\begin{align}\label{passive}
{\bm \theta}_1^{\star}= \arg \underset{ {\bm \theta}_1}{\text{max}}~\left|c({\bm \theta}_1)\right|
=e^{j \psi_1}\left( {\bar{\bm h}}_{I_1}\odot  {\bm a}_{I_1}( \vartheta_{I_1}, \varphi_{I_1}) \right)^*,
\end{align}
with $\psi_1 \in [0, 2\pi)$ denoting a common phase-shift of IRS~1 (to be specified later), and thus we define the optimal value as $c(\psi_1)\triangleq c({\bm \theta}_1^{\star})= e^{j \psi_1} \rho_{I_1} \delta_{I_1-G}M_1$. Accordingly, by substituting \eqref{passive} into \eqref{f1} with $c({\bm \theta}_1^{\star})= e^{j \psi_1} \rho_{I_1} \delta_{I_1-G}M_1$,
the objective function in \eqref{obj_P3} can be further written as
\begin{align}\label{optimal}
&\left\| {\bm f}_1({\bm \theta}_1) \right\|^2=\left\| \rho_{G} {\bm a}_G( \vartheta_G, \varphi_G)+e^{j \psi_1}\rho_{I_1} \delta_{I_1-G}M_1 {\bar{\bm h}}_{G} \right\|^2\notag\\
=&|\rho_{G}|^2 \left\|{\bm a}_G( \vartheta_G, \varphi_G)\right\|^2+ M_1^2|\rho_{I_1} \delta_{I_1-G}|^2\left\| {\bar{\bm h}}_{G}\right\|^2+2M_1\Re\Big(e^{j \psi_1} \underbrace{\rho^*_{G} \rho_{I_1} \delta_{I_1-G} {\bm a}^H_G( \vartheta_G, \varphi_G)  {\bar{\bm h}}_{G}}_{{\tilde \rho}_1\triangleq |{\tilde \rho}_1| e^{j {\tilde\psi_1}}}  \Big)\notag\\
=&N_1 |\rho_{G}|^2+ N_1 M_1^2|\rho_{I_1} \delta_{I_1-G}|^2+2M_1 |{\tilde \rho}_1|\cos \left(\psi_1+{\tilde\psi_1}\right)\notag\\
\stackrel{(a)}{ \le} &N_1 |\rho_{G}|^2+ N_1 M_1^2|\rho_{I_1} \delta_{I_1-G}|^2+2M_1 |{\tilde \rho}_1|=\gamma^\star_1,
\end{align}
where $\left\|{\bm a}_G( \vartheta_G, \varphi_G)\right\|^2=\left\| {\bar{\bm h}}_{G}\right\|^2=N_1$ for the array response at the GN, and
$(a)$ is due to $\cos \left(\psi_1+{\tilde\psi_1}\right)\le 1$ with the equality holding if and only if 
$\psi_1=-{\tilde\psi_1}$, which indicates that the active and passive beamforming are coherently combined at the GN side for maximizing $\left\|{\bm f}_1({\bm \theta}_1)\right\|^2$. As such, the optimal common phase-shift in \eqref{passive} to maximize the gain in \eqref{optimal} should be set as} 
\begin{align}\label{common_phase}
\psi_1^\star=-{\tilde\psi_1}=\angle \left(\rho_{G}\right) -\angle \left(\rho_{I_1}\right)- \angle \left(\delta_{I_1-G} {\bm a}^H_G( \vartheta_G, \varphi_G)  {\bar{\bm h}}_{G}\right).
\end{align}
\rev{Accordingly, the optimal solution to problem (P3) is derived in closed-form with the expressions given in \eqref{passive} and \eqref{common_phase}, which achieves the {\it globally} optimal value given in the right hand side of \eqref{optimal}.} As compared to the SDR algorithm in \cite{Wu2019TWC}, the closed-form passive beamforming design based on \eqref{passive} and \eqref{common_phase} has much lower complexity to compute and thus is more appealing to the high-mobility satellite communication.
\rev{Following the similar procedure, we can also derive the optimal closed-form solution to problem (P2) for the SAT side with $\mu=2$ (for which the details are omitted for brevity) and thereby the {\it globally} optimal value to problem (P1) is given as $\gamma^\star=\gamma^\star_1\cdot \gamma^\star_2$.}
In particular, it can be inferred from \eqref{optimal} that
if $M_1=M_2={\bar M}$ and $N_1=N_2={\bar N}$ for symmetric (active/passive) array deployment at the GN and SAT sides,
the IRS-aided satellite communication with two-sided cooperative IRSs has an effective channel gain with the maximum power scaling order of ${\cal O}\left({\bar N}^2 {\bar M}^4\right)$, which can effectively compensate for the severe path loss due to its long propagation distance.

\emph{Remark 1:} According to the decomposition structure shown in \eqref{Eff_Ch3} and \eqref{decouple}, \rev{the distributed local CSI at the GN and SAT sides is sufficient} for designing the active and passive beamforming $\left\{{\bm \theta}_1,{\bm w}_1\right\}$ and $\left\{{\bm \theta}_2,{\bm w}_2\right\}$, respectively, without loss of optimality.
Specifically, at the GN side, since the IRS~1$\rightarrow$GN channel ${{\bm H}}_{I_1-G}$ as well as its rank-one decomposition in \eqref{rank-one} can be determined offline based on the fixed geometric relationship,
we only need to acquire the local real-time CSI in terms of AoA pairs $\left\{( \vartheta_G, \varphi_G), ( \vartheta_{I_1}, \varphi_{I_1})\right\}$ and the phase-shift difference $\Delta_{\rho1} \triangleq\angle \left(\rho_{G}\right) -\angle \left(\rho_{I_1}\right)$ for the joint active and passive beamforming design $\left\{{\bm \theta}_1^{\star},{\bm w}_1^{\star}\right\}$ based on \eqref{active}, \eqref{passive}, and \eqref{common_phase}.
Similarly, at the SAT side, with the SAT$\rightarrow$IRS~2 channel ${{\bm H}}_{S-I_2}$ obtained offline,
we only need to acquire the local real-time CSI in terms of AoA pairs $\left\{( \vartheta_S, \varphi_S), ( \vartheta_{I_2}, \varphi_{I_2})\right\}$ and the phase-shift difference $\Delta_{\rho2} \triangleq\angle \left(\rho_{S}\right) -\angle \left(\rho_{I_2}\right)$ for jointly designing the optimal active and passive beamforming $\left\{{\bm \theta}_2^{\star},{\bm w}_2^{\star}\right\}$.

\section{Distributed Channel Estimation and Beam Tracking}\label{CH_Est}
\begin{figure}[!t]
	\centering
	\includegraphics[width=5.5in]{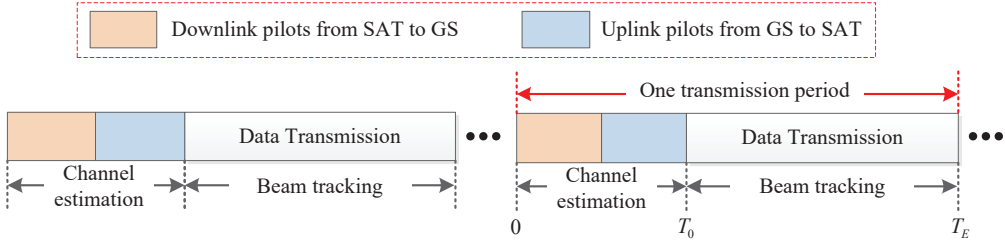}
	\setlength{\abovecaptionskip}{-3pt}
	\caption{Channel estimation and beam tracking protocol for the IRS-aided LEO satellite communication with two-sided IRSs.}
	\label{Protocol}
\end{figure}

As discussed in the previous section (cf. Remark 1), we only need to acquire the local real-time CSI at the GN and SAT sides for the active and passive beamforming design in a distributed manner.
Attentive to this, we propose a new and practical transmission protocol in this section to conduct distributed channel estimation with low training overhead as well as distributed beam tracking for the high-mobility satellite communication with two-sided IRSs efficiently.
The proposed protocol is shown in Fig.~\ref{Protocol}, which includes two main procedures that are conducted alternately as described below.
\begin{itemize}
	\item {\bf Channel estimation:} The SAT first transmits the downlink (beam) pilots, based on which the GN estimates its local CSI in terms of AoA pairs $\left\{( \vartheta_G^{[t]}, \varphi_G^{[t]}), ( \vartheta_{I_1}^{[t]}, \varphi_{I_1}^{[t]})\right\}$ and the phase-shift difference $\Delta_{\rho1}$. \rev{Based on the estimated local CSI at the GN side,} the GN and IRS~1 set the active and passive beamforming according to \eqref{active}, \eqref{passive}, and \eqref{common_phase}, and then the GN sends the uplink beam pilots towards the SAT for it to estimate its local CSI in terms of AoA pairs 
	$\left\{( \vartheta_S^{[t]}, \varphi_S^{[t]}), ( \vartheta_{I_2}^{[t]}, \varphi_{I_2}^{[t]})\right\}$ and the phase-shift difference $\Delta_{\rho2}$.
	\item  {\bf Beam Tracking:} After obtaining the estimated CSI by the end of each channel training period, both the GN and SAT will track/update their local CSI over time in a distributed manner by leveraging  the {\it prior} information on the altitude and velocity of the satellite.
	Based on the updated local CSI at the GN and SAT sides, the active and passive beamforming designs at the two sides are adjusted over time accordingly in a distributed manner for enhancing data transmission, until the next channel estimation/training is initiated. 
\end{itemize}
In the following two subsections, we elaborate the above two procedures, respectively, in detail.

\subsection{Distributed Channel Estimation}\label{CE}
In this section, we focus on the (initial) channel estimation in each transmission period (see Fig.~\ref{Protocol}) for the IRS-aided satellite communication with two-sided IRSs, where each channel training period is divided into two blocks (referred to as the downlink and uplink channel training), consisting of $I_D$ and $I_U$ pilot symbols, respectively.
Moreover, we consider a practically short duration for each channel training period, during which all the involved channels are assumed to remain approximately constant. As such, we drop the time index ${[t]}$ in this subsection without causing any confusion.
\subsubsection{Downlink Channel Training}
During the downlink channel training, the SAT transmits the downlink pilots to the GN for it to estimate its local CSI, where we fix the active and passive beamforming vectors $\{{\bm w}_2, {\bm \theta}_2\} $ at the SAT side\footnote{During the initial channel estimation without any reference CSI, the SAT can choose the pre-designed active and passive beamforming vectors $\{{\bm w}_2, {\bm \theta}_2\} $ (e.g., set as the traditional active array and passive reflect-array with fixed phase shifts \cite{Mousavi2008LowCost,ma2021development,imaz2020reflectarray,hum2013reconfigurable}) to broadcast the downlink pilots within its coverage range on the Earth; while for the subsequent (non-initial) channel estimation, the SAT can apply the latest updated local CSI for setting its active and passive beamforming vectors $\{{\bm w}_2^{\star}, {\bm \theta}_2^{\star}\} $ to beam the downlink pilots towards the designated GN for enhancing the channel estimation performance.}
 and dynamically tune the reflection vector of IRS~1, i.e., ${\bm \theta}_1^{(i)}$, over different pilot symbols to facilitate the channel estimation at the GN side.
Based on the channel model in \eqref{Eff_Ch3}, the received signal at the GN is expressed as
\begin{align}\label{GN_rec}
{\bm y}_{G}^{(i)}&=\left(\rho_{G} {\bm a}_G( \vartheta_G, \varphi_G)+\rho_{I_1} {{\bm H}}_{I_1-G} {\bm \Theta}_1^{(i)}  {\bm a}_{I_1}( \vartheta_{I_1}, \varphi_{I_1})\right)  {\bm f}^T_2({\bm \theta}_2) {\bm w}_2 x_{S}^{(i)}+{\bm v}_{G}^{(i)}\notag\\
&=\underbrace{\left[
{\bf I}_{N_1}, {{\bm H}}_{I_1-G} {\bm \Theta}_1^{(i)} \right]}_{{\bm G}^{(i)}}
\begin{bmatrix}
{\bar\rho}_{G} {\bm a}_G( \vartheta_G, \varphi_G) \\
{\bar\rho}_{I_1}{\bm a}_{I_1}( \vartheta_{I_1}, \varphi_{I_1})\end{bmatrix} +{\bm v}_{G}^{(i)}, \quad i=1,\ldots,I_D,
\end{align}
where $x_{S}^{(i)}$ represents the pilot symbol transmitted by the SAT which is simply set as $x_{S}^{(i)}=1$ for ease of exposition,
$I_D$ is the number of downlink pilot symbols received by the GN,
${\bm v}_{G}^{(i)}\sim {\mathcal N_c }({\bm 0}, \sigma^2{\bm I}_{N_1})$ is the additive white Gaussian noise (AWGN) vector at the GN with $\sigma^2$ being the normalized noise power,
and we define ${\bar\rho}_{G}=\rho_{G} {\bm f}^T_2({\bm \theta}_2) {\bm w}_2$ and ${\bar\rho}_{I_1}=\rho_{I_1} {\bm f}^T_2({\bm \theta}_2) {\bm w}_2$ as the effective path gains for notational simplicity.
By  stacking the received signal vectors $\left\{{\bm y}_{G}^{(i)}\right\}_{i=1}^{I_D}$ into ${\bm y}_{G}=\left[({\bm y}_{G}^{(1)})^T,\ldots, ({\bm y}_{G}^{(I_D)})^T\right]^T$, we have
\begin{align}\label{GN_rec2}
{\bm y}_{G}= \underbrace{\begin{bmatrix}
{\bm G}^{(1)}\\
\vdots\\
{\bm G}^{(I_D)}
\end{bmatrix}}_{{\bar{\bm G}}}
\begin{bmatrix}
{\bar\rho}_{G} {\bm a}_G( \vartheta_G, \varphi_G) \\
{\bar\rho}_{I_1}{\bm a}_{I_1}( \vartheta_{I_1}, \varphi_{I_1})\end{bmatrix} +
\underbrace{\begin{bmatrix}
	{\bm v}_{G}^{(1)}\\
	\vdots\\
	{\bm v}_{G}^{(I_D)}
	\end{bmatrix}}_{{\bar{\bm v}}_{G}},
\end{align}
where ${\bar{\bm G}}$ denotes the observation matrix at the GN and ${\bar{\bm v}}_{G}$ is the corresponding AWGN vector. Based on \eqref{GN_rec2}, the ML estimation of the AoA pairs and path gains at the GN side is given by
\begin{align}\label{ML_est}
\begin{Bmatrix}
{\hat{\bar\rho}}_{G}, ( {\hat\vartheta}_G, {\hat\varphi}_G) \\
{\hat{\bar\rho}}_{I_1}, ( {\hat\vartheta}_{I_1}, {\hat\varphi}_{I_1})\end{Bmatrix}=\arg ~~\min_{\scriptsize \begin{Bmatrix}
	{\bar\rho}_{G}, ( \vartheta_G, \varphi_G) \\
	{\bar\rho}_{I_1}, ( \vartheta_{I_1}, \varphi_{I_1})\end{Bmatrix}}~~~~\left\|{\bm y}_{G}-{\bar{\bm G}}\begin{bmatrix}
{\bar\rho}_{G} {\bm a}_G( \vartheta_G, \varphi_G) \\
{\bar\rho}_{I_1}{\bm a}_{I_1}( \vartheta_{I_1}, \varphi_{I_1})\end{bmatrix}\right\|^2.
\end{align}
However, the ML estimation in \eqref{ML_est} encounters considerably high computational complexity due to the joint search over $\left\{ {\bar\rho}_{G},( \vartheta_G, \varphi_G), {\bar\rho}_{I_1}, ( \vartheta_{I_1}, \varphi_{I_1})  \right\}$.
\rev{To tackle this issue and make it as simple as possible for practical implementation,} we further propose a low-complexity decoupled channel estimation scheme by first estimating the AoA pairs and then the path gains successively.
Specifically, based on the LS criterion, we left-multiply ${\bm y}_{G}$ in \eqref{GN_rec2} by ${\bar{\bm G}}^{\dagger}=\left({\bar{\bm G}}^H{\bar{\bm G}}\right)^{-1} {\bar{\bm G}}^H$ and thus we have 
\begin{align}\label{GN_rec3}
\begin{bmatrix}
{\bar{\bm y}}_{G} \\
{\bar{\bm y}}_{I_1}\end{bmatrix}\triangleq{\bar{\bm G}}^{\dagger}{\bm y}_{G}= 
\begin{bmatrix}
{\bar\rho}_{G} {\bm a}_G( \vartheta_G, \varphi_G) \\
{\bar\rho}_{I_1}{\bm a}_{I_1}( \vartheta_{I_1}, \varphi_{I_1})\end{bmatrix} +{\bar{\bm G}}^{\dagger}{\bar{\bm v}}_{G},
\end{align}
where ${\bar{\bm y}}_{G}\in {\mathbb{C}^{N_1\times 1 }}$ and ${\bar{\bm y}}_{I_1} \in {\mathbb{C}^{M_1\times 1 }} $.
According to \eqref{GN_rec3}, existing AoA estimation algorithms such as multiple
signal classification (MUSIC) can be applied to estimate AoA pairs $( \vartheta_G, \varphi_G)$ and $( \vartheta_{I_1}, \varphi_{I_1})$ in a decoupled manner based on ${\bar{\bm y}}_{G}$ and 
${\bar{\bm y}}_{I_1}$, respectively. Moreover, with the estimated AoA pairs $( {\hat\vartheta}_G, {\hat\varphi}_G)$ and $( {\hat\vartheta}_{I_1}, {\hat\varphi}_{I_1})$, the LS estimates of the path gains are given by
\begin{align}
{\hat{\bar\rho}}_{G}=\frac{{\bm a}^H_G( {\hat\vartheta}_G, {\hat\varphi}_G) {\bar{\bm y}}_{G}}{N_1}, \qquad
{\hat{\bar\rho}}_{I_1}=\frac{{\bm a}^H_{I_1}( {\hat\vartheta}_{I_1}, {\hat\varphi}_{I_1}) {\bar{\bm y}}_{I_1}}{M_1}.
\end{align}
Recall that ${\bar\rho}_{G}/{\bar\rho}_{I_1}=\rho_{G}/\rho_{I_1}$ and
the phase-shift difference is estimated as
\begin{align}\label{phase-shift}
{\hat \Delta}_{\rho1}=\angle \left({{\hat\rho}_{G}}\right) -\angle \left({\hat\rho}_{I_1}\right)=\angle \left({{\hat{\bar\rho}}_{G}}\right) -\angle \left({\hat{\bar\rho}}_{I_1}\right).
\end{align}
\rev{With the estimated CSI obtained in \eqref{ML_est}-\eqref{phase-shift} to replace the perfect CSI, we can design the active and passive beamforming vectors $\{{\bm w}^{\star}_1, {\bm \theta}^{\star}_1\} $ at the GN side according to \eqref{active}, \eqref{passive}, and \eqref{common_phase} for practical implementation.}\footnote{\rev{Note that the robust passive beamforming design highly depends on the adopted CSI error model, which, however, is beyond the scope of this paper and will be considered
		in our future work.}} 

\subsubsection{Uplink Channel Training}
During the uplink channel training, the GN transmits the uplink beam pilots to the SAT for it to estimate its local CSI, where we apply the previously optimized active and passive beamforming vectors $\{{\bm w}^{\star}_1, {\bm \theta}^{\star}_1\} $ at the GN side and dynamically tune the reflection vector of IRS~2, i.e., ${\bm \theta}_2^{(i)}$, over different pilot symbols to facilitate the channel estimation at the SAT side. \rev{In particular, by swapping the two sides of the GN and SAT and following the similar procedures in \eqref{GN_rec}-\eqref{phase-shift}, we can obtain the estimates of the AoA pairs and the phase-shift difference at the SAT side, based on which the active and passive beamforming vectors $\{{\bm w}^{\star}_2, {\bm \theta}^{\star}_2\} $ can be similarly designed at the SAT side (with the details omitted for brevity).}
\subsection{Distributed Beam Tracking}
During each transmission period as shown in Fig.~\ref{Protocol},
we let $( {\hat{\vartheta}}_X^{[T_0]}, {\hat{\varphi}}_X^{[T_0]})$ with $X \in \{G, I_1, S, I_2\}$ denote the initially estimated AoA pair of node $X$ at the beginning of each data transmission frame (or equivalently, by the end of each channel training period), where we let $T_0$ and $T_E$ denote the start and end time of one data transmission frame of interest.
During each data transmission frame, we propose to model/approximate the variation of the AoA pair $( {{\vartheta}}_X^{[t]}, {{\varphi}}_X^{[t]})$ (due to the relative motion between the GN and SAT sides) as a linear time-varying process.\footnote{Note that since the GN and IRS~1 have a fixed geometric relationship, the corresponding phase-shift difference ${\Delta}_{\rho1}=\angle \left({\bar\rho}_{G}\right) -\angle \left({\bar\rho}_{I_1}\right)$ at the GN side remains almost unchanged during each data transmission frame, regardless of the Doppler effect (induced by the high-mobility satellite) on both ${\bar\rho}_{G}$ and ${\bar\rho}_{I_1}$. This also holds for the corresponding phase-shift difference ${\Delta}_{\rho2}$ at the SAT side due to the fixed geometric relationship between the SAT and IRS~2, despite their high mobility.} \rev{Accordingly, the AoA information at the GN and SAT sides under the 3D Cartesian coordinate system shown in Fig.~\ref{system} can be predicted/tracked as follows.
\begin{align}\label{thetaX}
{\hat{\vartheta}}_X^{[t]}={\hat{\vartheta}}_X^{[T_0]}+\left(t-T_0\right) {\Delta{\vartheta}}_X, \qquad
{\hat{\varphi}}_X^{[t]}={\hat{\varphi}}_X^{[T_0]}+\left(t-T_0\right) {\Delta{\varphi}}_X,
\end{align}
with $t \in [T_0, T_E]$ and
$X \in \{G, I_1, S, I_2\}$, where ${\Delta{\vartheta}}_X$ and ${\Delta{\varphi}}_X$ denote the linear increments of the AoA pair over time due to the relative motion between the GN and SAT sides.}
Note that since the orbit of the satellite is known\cite{Chen2020Vision},
the above linear increments $\left\{{\Delta{\vartheta}}_X, {\Delta{\varphi}}_X\right\}$ can be easily determined by leveraging the {\it prior} information in terms of altitude and velocity of the satellite, as specified in the following.
 \begin{figure}[!t]
	\centering
	\includegraphics[width=5.5in]{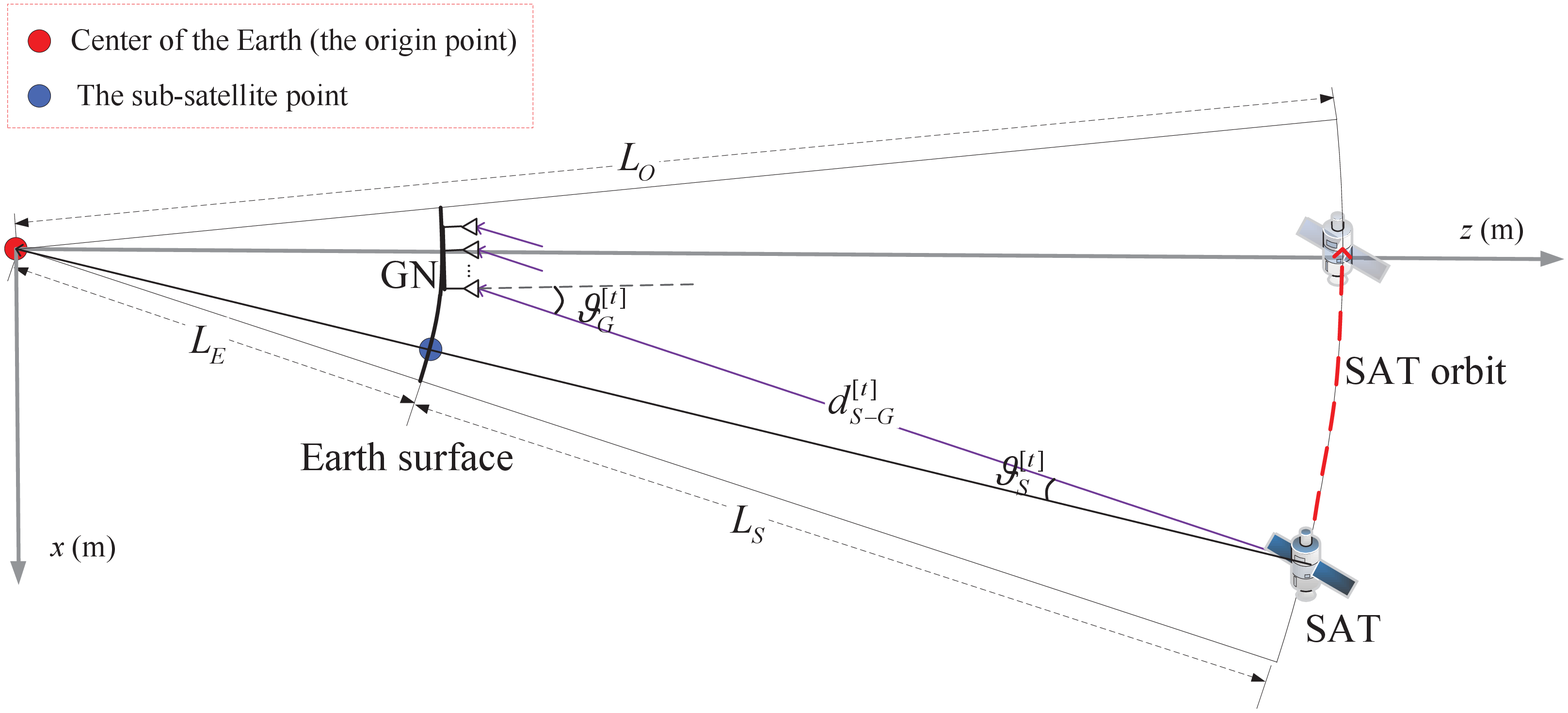}
	\setlength{\abovecaptionskip}{-3pt}
	\caption{An illustration for the geometric relationship between the GN and SAT (2D view).}
	\label{SAT_position}
\end{figure}

Let us consider an illustrative example for obtaining the linear increment ${\Delta{\vartheta}}_G$ in the considered IRS-aided satellite communication,
where the geometric relationship between the GN and SAT is shown in Fig.~\ref{SAT_position}.
Moreover, we consider that the GN and the satellite's orbit are \rev{within the same $x-z$ plane with the Earth center being the original point}, where we have $\varphi_G^{[t]}=\varphi_S^{[t]}=0$ for ease of illustration.\footnote{\rev{The results in this paper can be readily extended to the general 3D case by taking into account the geometric angle formed by the GN location and the satellite's orbit, for which we have $\varphi_G^{[t]}\ne 0$ and $\varphi_S^{[t]}\ne 0$ in general.}}
Furthermore, we assume that during each data transmission frame, the LEO satellite orbits the Earth at a constant speed of $V$. 
\rev{Based on the geometric relationship between the GN and SAT over time, we approximate the linear increment ${\Delta{\vartheta}}_G$ as follows.\footnote{\rev{Under the general 3D case, both the orbital speed $V$ and the average distance ${\bar d}_{S-G}$ need to be decomposed into two components projected onto the $x-z$ and $y-z$ planes, respectively.}}
\begin{align} 
{\Delta{\vartheta}}_G= \frac{1}{T_E-T_0}\left({{\vartheta}}_G^{[T_E]}-{{\vartheta}}_G^{[T_0]}\right)
\approx \pm \frac{V}{{\bar d}_{S-G}}
\end{align}
where ${\bar d}_{S-G}=\frac{1}{T_E-T_0}\int_{T_0}^{T_E}d_{S-G}^{[t]}~dt$ denotes the average distance between the SAT and GS during the data transmission frame of interest (which can be calculated by exploiting the known satellite's trajectory),
and the sign ``$\pm $" depends on the moving direction of the SAT.}
While other linear increments can also be similarly obtained, whose details are thus omitted for brevity.

\emph{Remark 2:} Based on the AoA information predicted/tracked in \eqref{thetaX} over time, the GN and SAT sides can dynamically adjust their active and passive beamforming vectors in real time according to \eqref{active}, \eqref{passive}, and \eqref{common_phase} during each data transmission frame to enhance the data transmission performance.
It is also noted that by the end of each data transmission frame, the CSI prediction in \eqref{thetaX} may cause some deviations from the actual CSI over time, due to the estimation errors and accumulative prediction errors. As such, periodic channel estimation (following each data transmission frame) is necessary to maintain the high effective channel gain for the IRS-aided satellite communication system with two-sided cooperative IRSs, as will be shown via simulations in the next section.

\section{Simulation Results}\label{Sim}
In this section, we present simulation results to examine the performance of the considered
IRS-aided LEO satellite communication with two-sided cooperative IRSs. Moreover, we also validate the effectiveness of the proposed joint active and passive beamforming design as well as channel estimation and beam tracking protocol via simulations.
We consider that the LEO satellite orbits the Earth at a fixed altitude of
$L_S=6\times 10^5$~m and at a constant speed of $V=7.5665\times 10^2$~m/s (which is determined by the gravitational constant and the mass of the Earth) for serving GNs in a periodic manner. Accordingly, given the Earth's radius $L_E=6.37\times 10^6$~m, the radius of the satellite's orbit is determined as $L_O=L_S+L_E=6.97\times 10^6$~m and thus the satellite's orbital period is given by $T_P=\frac{2 \pi L_O }{V}\approx 96$ minutes.\footnote{In practice, multiple LEO satellites in a sequel may be required to shorten the service time gap for a given GN.}
As illustrated in Fig.~\ref{SAT_position}, the plane formed by the satellite's orbit is considered as the reference $x-z$ plane, where the Earth center is the origin point.
\rev{Under the considered 3D Cartesian coordinate system,} we assume that the central (reference)
points of the GN and IRS~1 are located at the fixed positions of $(0, 0, L_E+100 )$~m and $(5, 0, L_E+95 )$~m on the Earth, respectively; while the central (reference) points of the SAT and IRS~2 at time $t$ are represented by $\left(L_O \sin\left(\frac{V}{L_O }t\right) , 0, L_O \cos\left(\frac{V}{L_O }t\right) \right)$~m and $\left(L_O \sin\left(\frac{V}{L_O }t\right)+3 , 0, L_O \cos\left(\frac{V}{L_O }t\right)+3 \right)$~m, respectively, with $0\le t \le T_P$ in one orbital period of interest.
In particular, under the considered setup, we have $\varphi_G^{[t]}=\varphi_S^{[t]}=\varphi_{I_1}^{[t]}=\varphi_{I_2}^{[t]}=0$ and thus we only need to focus on the real-time variations of AoAs/AoDs $\left\{ \vartheta_G^{[t]}, \vartheta_S^{[t]}, \vartheta_{I_1}^{[t]},\vartheta_{I_2}^{[t]}\right\} $ at the GN, SAT, IRS~1, and IRS~2, respectively, in our simulations.

On the other hand, the GN and SAT are equipped with UPAs that consist of $N_{1}=5\times5$ and $N_{2}=5\times5$ active antennas, respectively.
The reference path gain at the distance of $1$~m is set as $\beta=-30$~dB for all individual links.
For the purpose of exposition only, we assume that the GN and SAT have equal transmit power denoted by $P_T$ and the noise power at
the GN and SAT is set as $\sigma_N^2=-90$~dBm. Accordingly, the normalized noise power at the GN and SAT is given by $\sigma^2=\sigma_N^2/P_T$.
We assume that the satellite communication system operates at the very high frequency (VHF) of $150$ megahertz (MHz) with the
wavelength of $\lambda=2$ m.
Moreover, we set the equal antenna/element spacing at the GN, SAT, IRS~1, and IRS~2 as $\Delta_G=\Delta_S=\Delta_{I1}=\Delta_{I2}=\lambda/8=0.25$ m.

\subsection{Beamforming Performance Comparison Under Perfect CSI}
In this subsection, we study the beamforming performance under the assumption that the perfect real-time CSI on the LoS components of $\left\{{{{\bm H}}_{S-G}^{[t]}}, {{\bm H}}_{S-I_1}^{[t]}, {{\bm H}}_{I_2-G}^{[t]}, {{\bm H}}_{I_2-I_1}^{[t]}, {{\bm H}}_{I_1-G}, {{\bm H}}_{S-I_2} \right\}$ is available. 
For all the involved links, we consider the Rician fading channel model with the Rician factor denoted by $\kappa$, which will be specified later to study the effect of the NLoS channel components on the system performance.
\begin{figure}[!t]
	\centering
	\includegraphics[width=4.5in ]{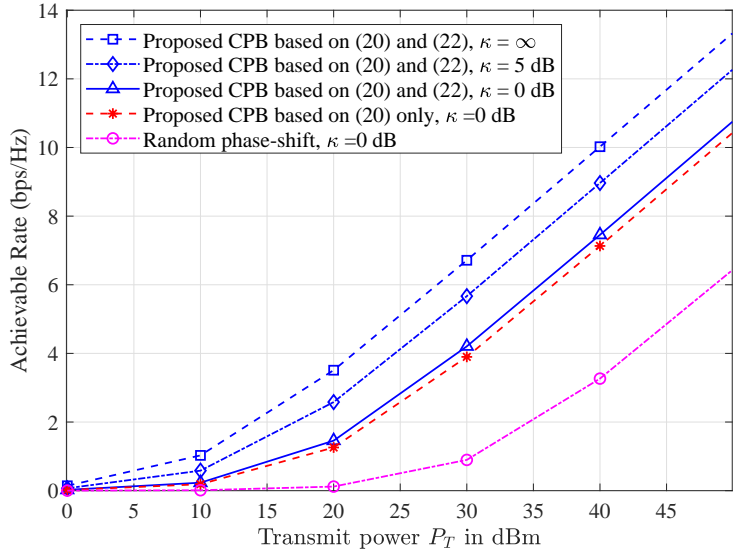}
	\setlength{\abovecaptionskip}{-5pt}
	\caption{Achievable rate versus transmit power $P_T$, with $M_1=M_2=500$.}
	\label{beamforming_power}
\end{figure}

In Fig.~\ref{beamforming_power}, we plot the achievable rate versus the transmit power $P_T$ for different passive beamforming designs and Rician factors in the considered IRS-aided satellite communication with two-sided IRSs.
As the Rician factor $\kappa$ decreases, it is observed that the proposed CPB design based on \eqref{passive} and \eqref{common_phase} suffers from a lower effective channel gain and thus its achievable rate decreases accordingly. This is expected since the increased power in the NLoS components will degrade the performance of the proposed CPB design based on \eqref{passive} and \eqref{common_phase} that caters for the LoS components only.
On the other hand, under the same Rician factor of $\kappa=0$~dB, by optimizing the common phase-shift $\psi_1$ for
aligning the active and passive beamforming, the proposed CPB design based on \eqref{passive} and \eqref{common_phase} achieves about $1$~dB power gain over that based on \eqref{passive} only.
Moreover, we consider the benchmark design where the reflection coefficients in ${\bm \theta}_\mu^{[t]}$ with $\mu\in \{1,2\}$ are
generated with random phase shifts following the uniform distribution within $[0, 2\pi)$ for comparison. It is observed that due to the large CPB gain over the LoS components,
the proposed CPB design significantly outperforms the random phase-shift benchmark.

For beamforming performance comparison given a budget on the total number of passive reflecting elements $M=M_1+M_2$, we consider the following baseline systems.
\begin{itemize}
	\item Satellite communication aided with {\bf SAT-side IRS only}: In this baseline, the communication between the GN and SAT is aided by the SAT-side IRS (i.e., IRS 2) only, which can also be considered as a special case of our proposed system with $M_1=0$ and $M_2=M$. \rev{According to Section~\ref{design}, the optimized IRS reflection vector at the SAT side is designed as ${\bm \theta}_2^{\star} 
	=e^{j \psi_2^{\star}}\left( {\bar{\bm h}}_{I_2}\odot  {\bm a}_{I_2}( \vartheta_{I_2}, \varphi_{I_2}) \right)^*$ with $\psi_2^{\star}$ being the optimal common phase-shift similar to \eqref{common_phase}.}
	\item Satellite communication aided with {\bf SAT-side reflect-array only}: In this baseline, the traditional passive reflect-array is mounted on the SAT side \rev{(i.e., on the reverse side of the satellite solar panels with $M_1=0$ and $M_2=M$) to introduce the fixed passive beamforming direction towards the ground at all time, for which we set the reflect-array vector as ${\bm \theta}_2
	=\left( {\bar{\bm h}}_{I_2}\odot  {\bm 1}_{M}\right)^*={\bar{\bm h}}_{I_2}^*$.} 
	\item \rev{Satellite communication aided with {\bf SAT-side reflect-array and GN-side IRS}: In this baseline, the communication between the GN and SAT is aided by the SAT-side reflect-array and the GN-side IRS with $M_1=M_2=M/2$, where we set the reflect-array vector at the SAT side as ${\bm \theta}_2
	=\left( {\bar{\bm h}}_{I_2}\odot  {\bm 1}_{M_2}\right)^*={\bar{\bm h}}_{I_2}^*$ and the optimized IRS reflection vector at the GN side as in \eqref{passive} and \eqref{common_phase}.}
	\item Satellite communication aided with {\bf GN-side IRS only}: In this baseline, the communication between the GN and SAT is aided by the GN-side IRS (i.e., IRS 1) only, which can also be considered as a special case of our proposed system with $M_1=M$ and $M_2=0$. \rev{Accordingly, the optimized IRS reflection vector at the GN side is designed as in \eqref{passive} and \eqref{common_phase}.}
	\item Satellite communication with {\bf no IRS/reflect-array}: In this baseline, only the active beamforming is performed at the GN and SAT sides, with no IRS/reflect-array at the two sides, i.e., $M_1=M_2=0$.
\end{itemize}

\begin{figure}[!t]
	\centering
	\includegraphics[width=4.5in ]{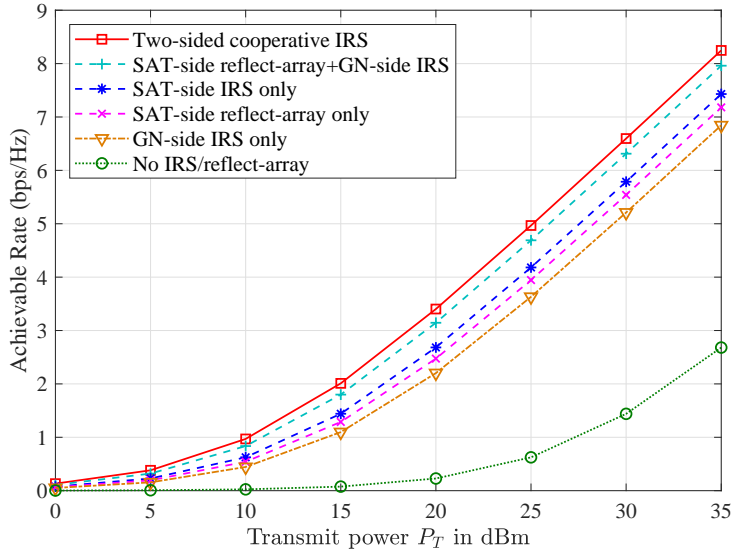}
	\setlength{\abovecaptionskip}{-5pt}
	\caption{\rev{Achievable rate versus transmit power $P_T$ under the practical discrete phase-shift model with $K=8$ and $M_1=M_2=500$.}}
	\label{beamforming_sch_power}
\end{figure}
\rev{Note that in practice, the phase shift of each IRS element can only take a
finite number of discrete values due to the hardware constraint \cite{you2019intelligent,wu2019beamforming}.
Specifically, let $K$ denote the number of phase shift levels and by uniformly quantizing the continuous
phase shift in the range of $[0,2\pi)$, the set of all possible
discrete phase shifts for each element can be represented by
${\cal F}\triangleq \left\{0, \Delta_\theta, \ldots,(K-1) \Delta_\theta \right\}$, where $\Delta_\theta=2\pi/K$.
As such, for any given continuous phase-shift design of each element (denoted by $\theta\in [0, 2\pi)$), we can directly quantize it to its nearest point in ${\cal F}$ to obtain the corresponding discrete phase-shift design for practical implementation, which is given by ${\bar \theta}=\arg~\min\limits_{{\bar \theta}\in {\cal F}} \left|{\bar \theta}- \theta\right|$. Under the discrete phase-shift model with $K=8$, we show in Fig.~\ref{beamforming_sch_power} the achievable rate versus the transmit power $P_T$ for different satellite communication systems at one particular time instant  of $t=10$~s.
It is observed that by exploiting the more pronounced passive beamforming gain, the proposed system aided with two-sided cooperative IRSs outperforms all the other baseline systems under the practical discrete phase-shift model, regardless of $P_T$.
Moreover, for both the one- and two-sided IRS deployment,
some performance loss can be observed by replacing the SAT-side IRS with the SAT-side reflect-array.
This is expected since the tunable elements in the SAT-side IRS can collaboratively achieve the fine-grained
passive beamforming gain.
On the other hand, the communication system aided with SAT-side IRS only achieves better performance than that with GN-side IRS only. 
This is because the shorter distance between the SAT and the SAT-side IRS leads to less product-distance path loss in the single-reflection link and thus the higher effective channel gain, as compared to the distance between the GN and the GN-side IRS with $d_{I_1-G}>d_{S-I_2}$ in our simulation setup. }

\begin{figure}[!t]
	\centering
	\includegraphics[width=4.5in ]{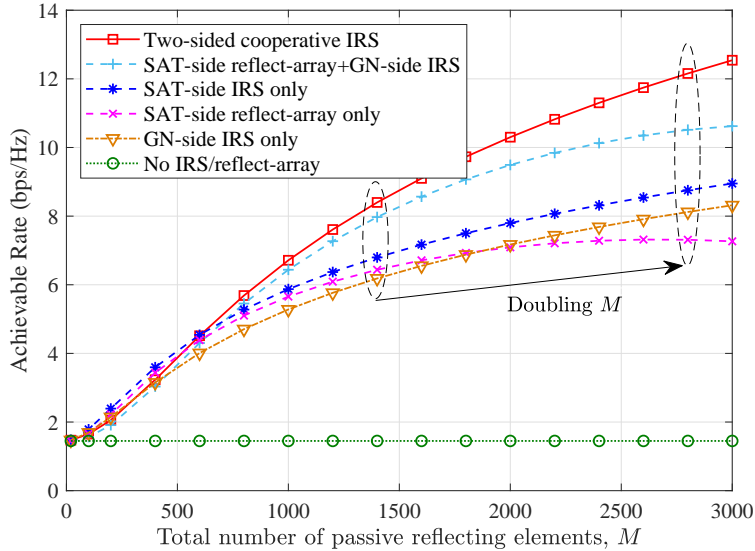}
	\setlength{\abovecaptionskip}{-5pt}
	\caption{Achievable rate versus the total number of passive reflecting elements $M$, with $P_T=30$~dBm and $\kappa=10$ dB.}
	\label{beamforming}
\end{figure}
In Fig.~\ref{beamforming}, we show the achievable rate versus the total number of passive reflecting elements $M$ for the proposed satellite communication aided with two-sided cooperative IRSs ($M_1=M_2=M/2$) against other baselines at one particular time instant  of $t=10$~s.
Several interesting observations
are made as follows. First, as the number of passive reflecting elements $M$ increases,
the achievable rates of all the satellite communication systems (except the one with no IRS/reflect-array) increase due to more signal reflection power. Second, one can observe that by doubling $M$ from $1,400$ to $2,800$, the achievable rate of the proposed communication system aided with two-sided cooperative IRSs increases about $\log_2(2^4)=4$~bps/Hz; whereas those with one-sided IRS (i.e., SAT-side IRS only or GN-side IRS only) only increases about $\log_2(2^2)=2$~bps/Hz.
This asymptotic performance gain is expected since their resultant effective channel gains have different maximum scaling orders, i.e., ${\cal O}\left(M^4\right)$ versus ${\cal O}\left(M^2\right)$ with increasing $M$ under the LoS-dominant channel condition.
Third, as compared to the communication system aided with SAT-side reflect-array that fixes the passive beamforming direction towards the ground (which may deviate from the designated GN as illustrated in Fig.~\ref{SAT_position}), the one aided with SAT-side IRS only can flexibly adjust the passive beamforming direction towards the designated GN more accurately, thus leading to a higher passive beamforming gain as well as higher achievable rate.
Finally, owing to the CPB gain with the coherent channel combining, the proposed system aided with two-sided cooperative IRSs achieves much better performance than all the other baseline systems and the performance gain becomes even larger by further increasing $M$.

\subsection{Performance Comparison for Channel Estimation/Beam Tracking}
In this subsection, we examine the performance of our proposed channel estimation and beam tracking protocol for the satellite communication aided with two-sided cooperative IRSs.
Note that at the time of submitting this work, there is no other work in the literature on designing the channel estimation and beam tracking for the IRS-aided satellite communication.
As such, for performance comparison, we consider the benchmark protocol where the distributed channel estimation scheme proposed in Section~\ref{CE} is applied to acquire the required local CSI at the GN and SAT sides during each channel training period, based on which the joint active and passive beamforming is designed according to Section~\ref{design} and then applied for the subsequent data transmission frame shown in Fig.~\ref{Protocol} (i.e., no additional beam tracking is implemented).
\begin{figure}[!t]
	\centering
	\includegraphics[width=4.5in ]{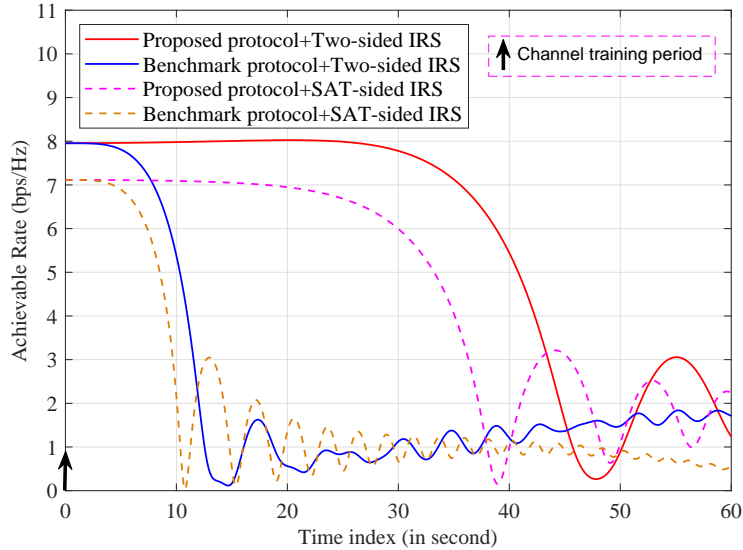}
	\setlength{\abovecaptionskip}{-5pt}
	\caption{\rev{Achievable rate versus time index, with $P_T=30$~dBm, $\kappa=10$ dB, and $M=1000$.}}
	\label{ch_tracking_long}
\end{figure}

\rev{In Fig.~\ref{ch_tracking_long}, we plot the achievable rate versus time index of the proposed and benchmark protocols for the two-sided IRS system (with $M_1=M_2=500$) and the SAT-side IRS only baseline system (with $M_1=0$ and $M_2=1000$).}
It is observed that the achievable rate of the benchmark protocol dramatically decreases over time with $t>5$~s and fluctuates at the rate of about $1$ bps/Hz with $t>15$~s. This can be explained by the fact that with the fixed active and passive beamforming design based on the estimated CSI at $t=0$, the effective channel gain will decrease due to the deviation from the actual beam direction over time in the high-mobility satellite communication.
In contrast, by dynamically adjusting the active and passive beamforming design according to the AoA prediction model in \eqref{thetaX}, the proposed protocol can well track the beam direction and thus achieves a high and stable effective channel gain (achievable rate) over a relatively longer duration, e.g., $0<t<30$~s.
Moreover, it is observed that the achievable rate of the proposed protocol also decreases over time when $t>30$~s, which is due to the estimation error and accumulative prediction error that cause the deviation from the actual beam direction over time.
Nevertheless, owing to the effective beam tracking, the proposed protocol has a much slower declining rate
than the benchmark protocol in terms of achievable rate over time.

\begin{figure}[!t]
	\centering
	\includegraphics[width=4.5in ]{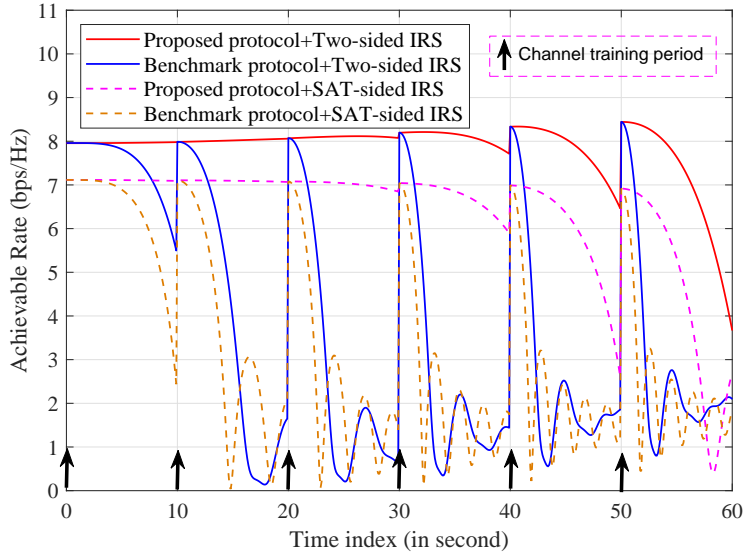}
	\setlength{\abovecaptionskip}{-5pt}
	\caption{\rev{Achievable rate versus time index, where each data transmission frame has the duration of $T_E-T_0=10$~s, $P_T=30$~dBm, $\kappa=10$ dB, and $M=1000$.}}
	\label{ch_tracking_short}
\end{figure}

In Fig.~\ref{ch_tracking_short}, we plot the achievable rate versus time index of the proposed and benchmark protocols using the periodic channel estimation as illustrated in Fig.~\ref{Protocol}, where each data transmission frame has the duration of $T_E-T_0=10$~s.
It is observed that the declining rates of both the proposed and benchmark protocols increase over time.
This is due to the fact that the smaller elevation angle from the GN to the SAT will result in  faster variations of the beam directions at both GN and SAT sides, thus causing a larger deviation over time.
Specifically, in our simulation setup, the SAT is directly above the GN at $t=0$ with the highest elevation angle; while the elevation angle from the GN to the SAT becomes smaller as $t$ increases.
\rev{On the other hand, by comparing with Fig.~\ref{ch_tracking_long}, it is observed that the periodic channel estimation in Fig.~\ref{ch_tracking_short} helps to maintain a higher effective channel gain (achievable rate) for the two-sided IRS system and the SAT-sided IRS only baseline system, by periodically correcting the deviation of the beam directions over time.}
Moreover, with the effective beam tracking based on the AoA prediction model in \eqref{thetaX}, the proposed protocol achieves a much higher and more stable achievable rate than the benchmark protocol using the fixed beam for each data transmission frame (i.e., no additional beam tracking is implemented).

\section{Conclusions}\label{conlusion}
In this paper, we proposed a new IRS-aided LEO satellite communication system with two-sided cooperative IRSs.
We formulated the joint active and passive beamforming design problem to maximize the overall channel gain between the SAT and GN. 
By decomposing the high-dimensional SAT-GN channel matrix into the outer product of two low-dimensional vectors, we decoupled this non-convex optimization problem into two simpler subproblems that were then solved in closed-form at the SAT and GN sides, respectively.
Moreover, based on the decomposed SAT-GN channel, we proposed a practical transmission protocol to conduct channel estimation and beam tracking at the SAT and GN sides efficiently in a distributed manner, which was shown to be able to achieve a high and stable effective channel gain for the high-mobility LEO satellite communication.
 Simulation results demonstrated the substantial performance gains achieved by the new double-IRS aided LEO satellite communication system under the proposed beamforming design and transmission protocol, as compared to various baseline systems.
\ifCLASSOPTIONcaptionsoff
  \newpage
\fi

\bibliographystyle{IEEEtran}
\bibliography{IRS_SAT}

\end{document}